\documentclass[reprint, aps, pra, twocolumn, nofootinbib, longbibliography]{revtex4-1}

\usepackage[utf8]{inputenc}
\usepackage{graphicx}
\usepackage{units}
\usepackage{color}
\usepackage[pdftex, colorlinks=true, linkcolor=myblue, citecolor=myblue,
urlcolor=myblue]{hyperref}
\usepackage{times}
\usepackage{bbold}

\definecolor{myblue}{rgb}{0,0,1}

\usepackage{graphicx}
\usepackage{amsmath}
\usepackage{amssymb}

\begin{document}

\title{Spontaneous orbital magnetization of mesoscopic dipole dimers}

\author{Ga\"etan J.\ Percebois}
\affiliation{Universit\'e de Strasbourg, CNRS, Institut de Physique et Chimie des Mat\'eriaux de Strasbourg,
UMR 7504, F-67000 Strasbourg, France}

\author{Dietmar Weinmann}
\affiliation{Universit\'e de Strasbourg, CNRS, Institut de Physique et Chimie des Mat\'eriaux de Strasbourg,
UMR 7504, F-67000 Strasbourg, France}

\author{Rodolfo A.\ Jalabert}
\affiliation{Universit\'e de Strasbourg, CNRS, Institut de Physique et Chimie des Mat\'eriaux de Strasbourg,
UMR 7504, F-67000 Strasbourg, France}

\author{Guillaume Weick}
\affiliation{Universit\'e de Strasbourg, CNRS, Institut de Physique et Chimie des Mat\'eriaux de Strasbourg,
UMR 7504, F-67000 Strasbourg, France}

\begin{abstract}
Ensembles of gold nanoparticles 
present a magnetic behavior which is at odds with the weakly diamagnetic response of bulk gold. 
In particular, an unusual ferromagnetic order has been unveiled by several experiments. 
Here we investigate if the combined effect of orbital magnetism of conduction electrons and interparticle dipolar interaction can lead to magnetic ordering.
Using different model systems of interacting mesoscopic magnetic dipoles, together with a microscopic description of the electron dynamics within the nanoparticles, we find that a spontaneous magnetic moment may arise in dimers of metallic nanoparticles when the latter are characterized by a large orbital paramagnetic susceptibility. 
\end{abstract}

\maketitle

\section{Introduction}

It is well established that bulk gold, when subject to an external magnetic field, 
has a diamagnetic behavior. Such a magnetic response is the result of orbital motion, with the largest contribution arising from core electrons 
(Langevin--Larmor diamagnetism), and a small component associated with conduction electrons (Landau diamagnetism), together with non-negligible spin (Pauli) and orbital (van Vleck) paramagnetic contributions 
 \cite{suzuk12_PRL}.

Numerous puzzling experiments have reported over the last two decades that assemblies of gold nanoparticles 
encapsulated with organic ligands can present either a
paramagnetic behavior \cite{hori99_JPA, nakae00_PhysicaB, hori04_PRB, yamam04_PRL, guerr08, guerr08_Nanotechnology, barto12_PRL, agrac17_ACSOmega}, 
a diamagnetic response larger than the one of the bulk \cite{cresp04_PRL, dutta07_APL, guerr08_Nanotechnology, rhee13_PRL, hori04_PRB}, 
or, even more surprisingly, a ferromagnetic instability \cite{cresp04_PRL, cresp06_PRL, dutta07_APL, donni07_AdvMater, garit08_NL, guerr08_Nanotechnology, guerr08, venta09, donni10_SM, maitr11_CPC, agrac17_ACSOmega, grege12_CPC}.\footnote{For a  review, see Ref.\ \cite{nealo12_Nanoscale}, which discusses the magnetic behavior of gold nanoparticle ensembles, which is quite variable among different experiments.}

Recently, the paramagnetic behavior of relatively dilute samples of noninteracting gold nanoparticles has been 
theoretically elucidated in terms of orbital magnetism of the confined conduction electrons \cite{gomez18_PRB}. Such an effect is a purely quantum-mechanical phenomenon, 
which in the bulk gives rise to the Landau diamagnetic susceptibility \cite{landau_statphys}
\begin{equation}
\label{eq:chi_L}
\chi_\mathrm{L}=-\frac{e^2 k_\mathrm{F}}{12\pi^2m_*c^2}.
\end{equation} 
Here, $e$ is the elementary charge, $m_*$ is the effective mass of the conduction electrons,\footnote{Here and in what follows, we identify the bare electron mass $m_\mathrm{e}$ and the effective mass $m_*$, as they have similar values for gold ($m_*\simeq1.1\,m_\mathrm{e}$).} $k_\mathrm{F}$ is the Fermi wave vector, and 
$c$ is the speed of light in vacuum.\footnote{Throughout this paper we use cgs units. Note that the (dimensionless) magnetic susceptibilities in SI and cgs units, $\chi_\mathrm{SI}$ and $\chi$, respectively, are connected through the relation 
$\chi_\mathrm{SI}=4\pi\chi$.} 
While in typical bulk metals 
$\chi_\mathrm{L}$ represents a minute contribution to the overall magnetic response (e.g., for gold $\chi_\mathrm{L}=-2.9\times10^{-7}$ while $\chi_\mathrm{bulk}=-2.7\times10^{-6}$ \cite{suzuk12_PRL}), the situation is totally different for constrained 
geometries \cite{richt96_PhysRep}. Indeed, in particles with nanometric dimensions,
the quantization of energy levels can lead to a very large orbital response with a 
zero-field susceptibility (ZFS) $|\chi|\gg|\chi_\mathrm{L}|$ that can be either paramagnetic ($\chi>0$) or diamagnetic ($\chi<0$), 
depending on the size of the individual nanoparticle \cite{ruite91_PRL, ruite93_MPLB, leuwe93_PhD, gomez18_PRB}. 
Moreover, the orbital response is temperature dependent, and paramagnetic peaks appear, whose height increases for decreasing temperature $T$.
When an average over nanoparticle sizes is performed, and when the dipolar interactions between each orbital magnetic moment are neglected, 
the response of the ensemble is paramagnetic \cite{gomez18_PRB}, in rather good quantitative agreement with the experimental data of Refs.\ \cite{hori99_JPA, nakae00_PhysicaB, hori04_PRB}.

The large diamagnetic response measured in certain samples, and in particular in the experiment of Ref.\ \cite{rhee13_PRL}, 
has been tentatively accounted for by Imry \cite{imry15_PRB} in terms of Aslamazov--Larkin superconducting fluctuations \cite{aslam75_JETP}
that persist at temperatures way above the critical one. However, the proposal of Ref.~\cite{imry15_PRB} is one to two orders of magnitude smaller than that measured in Ref.~\cite{rhee13_PRL}. An alternative theoretical proposal by Murzaliev, Titov, and Katsnelson \cite{murza19_PRB} invoked the spin--orbit coupling, which is important for gold atoms, as a mechanism turning the ensemble-averaged paramagnetic response into a diamagnetic one. 
However, it has been recently shown \cite{gomez21_preprint} that the extrinsic spin--orbit coupling due to the discontinuity of the electrostatic potential at the nanoparticle surface, as well as other relativistic 
and geometric effects, only lead to a small correction to the 
individual-particle ZFS calculated in Ref.~\cite{gomez18_PRB}. 

There have been several attempts to interpret the ferromagnetic response
measured in certain samples invoking
the formation of covalent bonds between the atoms residing at the 
surface of the nanoparticles and the ligands around it \cite{cresp04_PRL}, 
the Fermi-hole effect involving the surface atoms alone \cite{hori04_PRB, yamam04_PRL}, or 
electronic orbits circling around single domains of ligands \cite{herna06_PRL}. However, these interpretations have been 
ruled out by later experiments \cite{grege12_CPC, nealo12_Nanoscale}, and a mechanism explaining the observed ferromagnetic instability is still highly sought after.

Motivated by this long-standing puzzle of condensed matter physics, here we investigate if the large orbital magnetism of individual 
nanoparticles, as calculated in Ref.~\cite{gomez18_PRB}, together with the magnetic dipole--dipole interaction between the  nanoparticles, can be at the origin of magnetic order and the emergence of a large total magnetic moment. Indeed, since each nanoparticle can carry a huge magnetic moment 
at relatively weak applied fields (typically up to three orders of magnitude
larger than the Bohr magneton $\mu_\mathrm{B}=e\hbar/2m_\mathrm{e}c$, even at room temperature), one cannot exclude that interparticle interactions may
play a role for the magnetic properties of the samples. 

Treating the long-range and anisotropic dipolar interactions of permanent magnetic moments in a lattice constitutes a formidable computational task that can result in ferromagnetic transitions in particular geometries \cite{polit02_PRB, varon13_SR, alkad17_PRB, galli20_PRX}. The problem of
a macroscopic and disordered ensemble of metallic nanoparticles 
(as it is most likely the case in the experiments of Refs.~\cite{hori99_JPA, nakae00_PhysicaB, hori04_PRB, yamam04_PRL,  guerr08_Nanotechnology, barto12_PRL, agrac17_ACSOmega, cresp04_PRL, dutta07_APL, rhee13_PRL, cresp06_PRL, donni07_AdvMater, garit08_NL, guerr08, venta09, donni10_SM, maitr11_CPC, grege12_CPC}) is even more challenging because the magnetic moments do not have a uniform magnitude, but they result from the response to the local field according to a highly fluctuating susceptibility.

Instead of tackling the precise experimental situation (many details of which are unknown), in this paper we investigate the magnetic instability of 
a dimer of nanoparticles, which constitutes the building block of any realistic sample.
Using a microscopic description of the conduction electron dynamics within each individual nanoparticle we show that, surprisingly, 
a magnetization with aligned magnetic moments can appear at very low temperature when the individual nanoparticles composing the dimer are both paramagnetic.

Our paper is outlined as follows: In Sec.~\ref{sec:model} we briefly present our model of 
a metallic nanoparticle dimer interacting through the long-range magnetic dipolar coupling. 
In Sec.~\ref{sec:linear} we first consider the case of a linear orbital response which is valid for weak effective magnetic fields
and find the onset of a magnetic order with aligned (anti-aligned) magnetic moments, 
when both nanoparticles are paramagnetic (diamagnetic). 
Since the above-mentioned linear-response approach results in an unphysical magnetization above a critical value of the ZFS, 
in Sec.~\ref{sec:toy} we remedy for this issue by considering a model in which the orbital response of the nanoparticles saturates at large magnetic fields as typically occurs, 
and thus find that both ordered magnetic phases are stable. 
We then adopt in Sec.~\ref{sec:micro} the microscopic quantum-mechanical model developed in Refs.~\cite{gomez18_PRB,ruite91_PRL, leuwe93_PhD} and demonstrate that a magnetic order with aligned magnetic moments can be reached at very low temperature, while the anti-aligned configuration remains elusive. We conclude in Sec.~\ref{sec:ccl}. 

In Appendix \ref{app:PRB18} we briefly describe the microscopic model used in Sec.~\ref{sec:micro}
to characterize the magnetic response of an individual nanoparticle, signaling the different behavior that can be obtained when varying the temperature or the nanoparticle size.
In order to support the approach that disregards the thermal fluctuations of the magnetization of the nanoparticles, we present  in Appendix \ref{app:dimer} a model calculation describing the equilibrium properties of a mesoscopic dimer.  In Appendix \ref{app:chain} we extend some of our results to a chain of nanoparticles, and show that the mutual dipolar interactions between the magnetic moments increase the temperature below which magnetic order may arise.

\section{Nanoparticle dimer modeling}
\label{sec:model}

\begin{figure}[tb]
\begin{center}
\includegraphics[width=.9\linewidth]{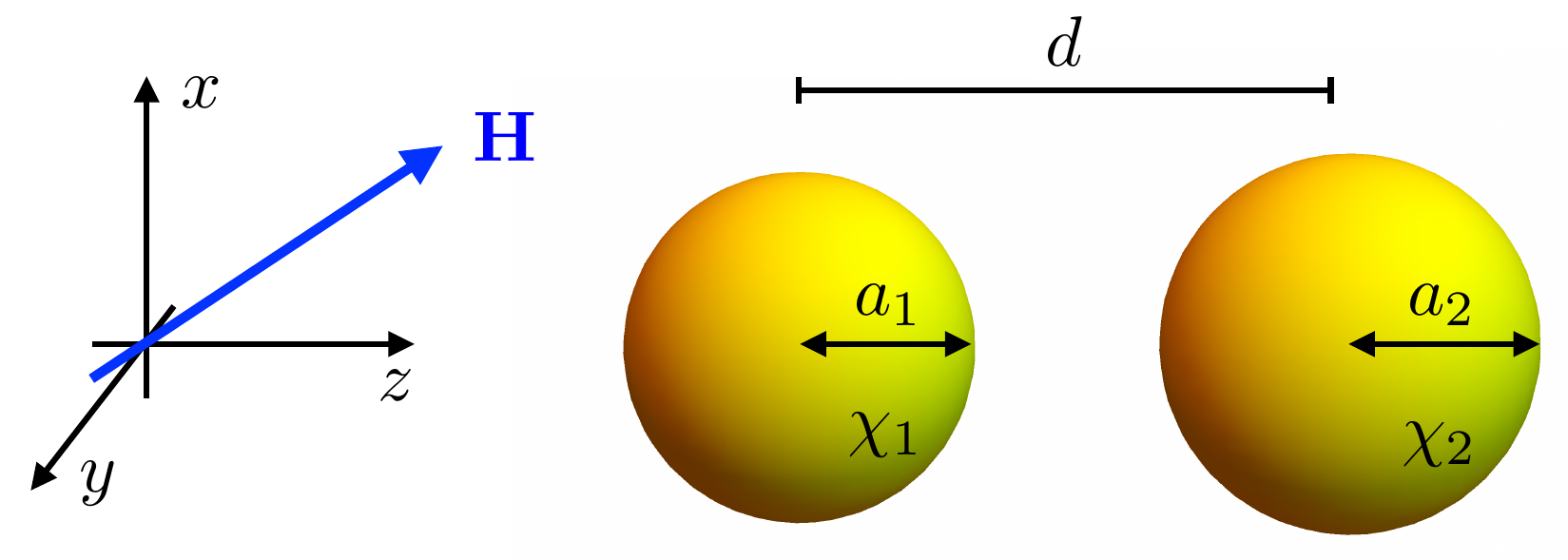}
\caption{\label{fig:sketch_dimer}%
Sketch of a dimer of spherical metallic nanoparticles of radii $a_1$ and $a_2$ spaced by a center-to-center distance $d$ and with, respectively, zero-field orbital susceptibilities $\chi_1$ and $\chi_2$, under the influence of an applied magnetic field $\mathbf{H}$.}
\end{center}
\end{figure}

We consider a dimer of spherical metallic, nonmagnetic nanoparticles
with radii $a_i$ ($i=1,2$). The dimer is aligned along the $z$ axis, and 
we call $d$ the center-to-center interparticle distance  (see Fig.~\ref{fig:sketch_dimer}). 
Each individual nanoparticle is assumed to carry a magnetic moment 
$\boldsymbol{\mathcal{M}}_i$ due to the orbital response to
the effective field $\mathbf{H}_i$ seen by nanoparticle $i$. 
The functional form of 
\begin{equation}
\label{eq:M_i}
\boldsymbol{\mathcal{M}}_i=\boldsymbol{\mathcal{M}}_i(\mathbf{H}_i, a_i, T)
\end{equation}
depends on the model adopted
to describe the electron dynamics within each nanoparticle, one example of it being given in Appendix \ref{app:PRB18}. This choice is crucial and will be thoroughly discussed in the sequel.
The total magnetic moment per particle (TMMPP) of the dimer reads
$\mathbf{m}=\sum_{i=1}^{2}\boldsymbol{\mathcal{M}}_i/{2}$.
The effective magnetic field
\begin{equation}
\label{eq:H_i}
\mathbf{H}_i=\mathbf{H}+\sum_{\substack{j=1\\(j\neq i)}}^2
\frac{3\hat{\mathbf{z}}\left(\hat{\mathbf{z}}\cdot\boldsymbol{\mathcal{M}}_j\right)-
\boldsymbol{\mathcal{M}}_j}{d^3}
\end{equation}
acting on the $i$th nanoparticle is given by 
the external applied field $\mathbf{H}$ and the contribution generated by the other 
nanoparticle through the magnetic dipole--dipole interaction.\footnote{Note that we consider interparticle separation distances $d\gtrsim3a$ so that we can disregard
higher multipolar terms in the interaction. Such a criterion was numerically verified in Ref.~\cite{park04_PRB} for oscillating electric dipoles in the 
context of localized surface plasmon resonances and used in Refs.~\cite{brand15_PRB, brand16_PRB, downi17_PRB} when, similarly to this work, a nanoparticle dimer was considered as the building block of larger assemblies. 
The induced field in Eq.~\eqref{eq:H_i} is evaluated at the nanoparticle center, since that value corresponds to 
the averaged field over its spherical volume \cite{Jackson}.}
Here and in the sequel of the paper, hats designate unit vectors. 

Expressions \eqref{eq:M_i} and \eqref{eq:H_i} thus represent a system of equations for the expectation values of the magnetizations $\boldsymbol{\mathcal{M}}_i$ that must be solved 
self-consistently. In such an approach, one disregards fluctuations of the magnetization around its thermal expectation value, assuming that the latter is the result of a large number of contributions, consistent with the large magnetizations of hundreds of Bohr magnetons $\mu_\mathrm{B}$ that can occur in an individual nanoparticle \cite{gomez18_PRB} (see also Appendix \ref{app:PRB18}, and in particular Fig.~\ref{fig:M_1NP}). A heuristic model for the magnetic behavior of two paramagnetic systems including thermal fluctuations presented in Appendix \ref{app:dimer} points to the validity of the approach based on Eqs.\ \eqref{eq:M_i} and \eqref{eq:H_i}.

\begin{figure*}[tb]
\begin{center}
\includegraphics[width=.65\linewidth]{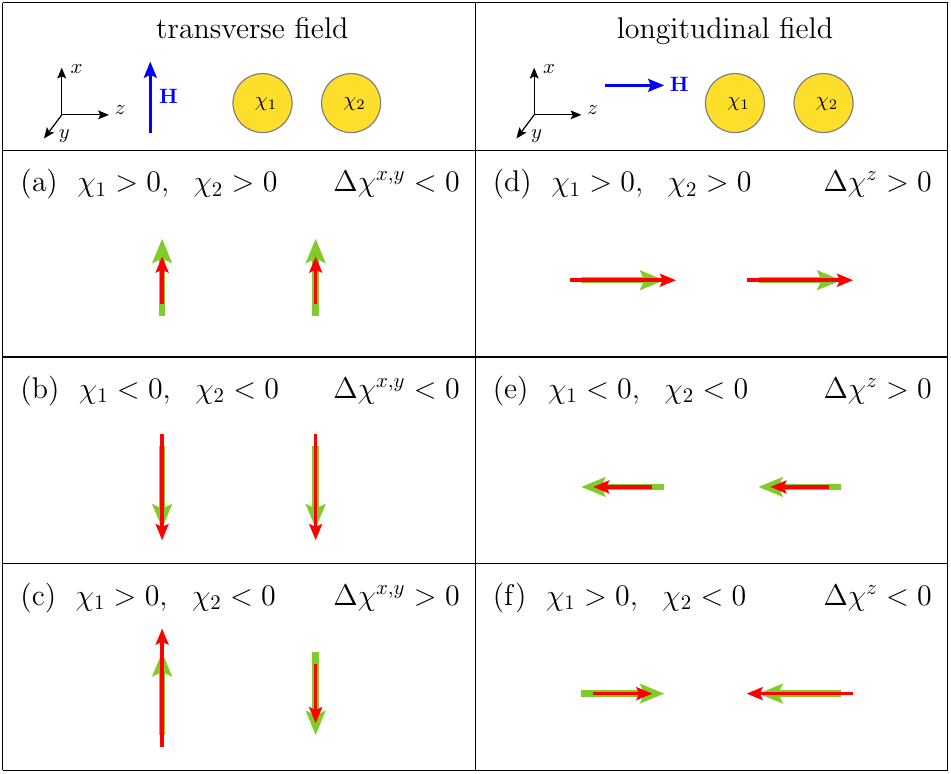}
\caption{\label{fig:interaction_dimer}%
Sketch of the magnetic response of the dimer with (red arrows) and without (green arrows)
dipolar interactions for (a)--(c) transverse and (d)--(f) longitudinal applied field $\mathbf{H}$. 
An arrow represents the magnetic moment of the $i$th nanoparticle pointing in the direction of $\boldsymbol{\mathcal{M}}_i$, 
and its length is proportional to $|\boldsymbol{\mathcal{M}}_i|$.}
\end{center}
\end{figure*}

For notational simplicity 
we will assume in what follows that the two nanoparticles have nearly identical sizes, $a_1\approx a_2=a$, but they still 
may have different ZFSs $\chi_1$ and $\chi_2$. Indeed, as we show in Appendix \ref{app:PRB18}, a difference $|a_1-a_2|$ of the order of $k_\mathrm{F}^{-1}\sim\unit[1]{\AA}$ may induce a radical change in the magnetic susceptibility.

\section{Linear response approach}
\label{sec:linear}

It is first instructive to consider that each nanoparticle remains in the linear regime, so that the magnetic moment 
of the $i$th nanoparticle is given by 
\begin{equation}
\label{eq:M_i_linear}
\boldsymbol{\mathcal{M}}_i=\mathcal{V}\chi_i\mathbf{H}_i ,
\end{equation}
with $\mathcal{V}=4\pi a^3/3$
the nanoparticle volume and $\mathbf{H}_i$ the effective magnetic field \eqref{eq:H_i}.
It is then straightforward to solve the system of equations \eqref{eq:H_i} and \eqref{eq:M_i_linear}, and we find
for the TMMPP the expression 
$\mathbf{m}=\mathcal{V}\overset{\leftrightarrow}{\boldsymbol{\chi}}\mathbf{H}$, where the diagonal 
zero-field
susceptibility 
tensor is given by
\begin{equation}
\overset{\leftrightarrow}{\boldsymbol{\chi}}=
\begin{pmatrix}
\chi^{xx} & 0 & 0 \\
0 & \chi^{yy} & 0 \\
0 & 0 & \chi^{zz}
\end{pmatrix},
\end{equation}
with 
\begin{equation}
\label{eq:chi_dimer}
\chi^{\sigma\sigma}= \frac 12\left( 
\frac{\chi_1+\chi_2-\eta_\sigma\frac{8\pi}{3}\chi_1\chi_2\left(\frac ad\right)^3}
{1-\eta_\sigma^2\frac{16\pi^2}{9}\chi_1\chi_2\left(\frac ad\right)^6}
\right), 
 \qquad \sigma=x,y,z.
\end{equation}
Here, $\eta_\sigma=1$ for the transverse directions ($\sigma=x,y$) and $\eta_\sigma=-2$ for the longitudinal one ($\sigma=z$). 

Let us now discuss the result of Eq.~\eqref{eq:chi_dimer} as a function of the interparticle distance $d$. 
Obviously, for a very large interparticle distance 
($d\gg a$), where the dipolar interaction, which scales as $1/d^3$, becomes negligible, the ZFS of the dimer corresponds to the average of the ZFSs of the two nanoparticles, i.e., 
$\lim_{d\to\infty}\chi^{\sigma\sigma}=\overline{\chi}$ with 
$\overline{\chi}=(\chi_1+\chi_2)/2$. The magnetic configuration of the noninteracting dimer is depicted by green arrows in Fig.~\ref{fig:interaction_dimer} for transverse (left column) and longitudinal (right column) applied fields $\mathbf{H}$ and for the three cases of interest 
$\{\chi_1>0, \chi_2>0\}$ [panels (a) and (d)], $\{\chi_1<0, \chi_2<0\}$ [panels (b) and (e)], and 
$\{\chi_1>0, \chi_2<0\}$ [panels (c) and (f)].\footnote{Obviously, the case $\{\chi_1<0, \chi_2>0\}$ 
is similar to $\{\chi_1>0, \chi_2<0\}$.} While in the two first cases the magnetic response of the dimer follows that of the individual nanoparticles, the response $\overline{\chi}$ in the latter case depends on the relative importance of $\chi_1$ and $\chi_2$.

For shorter interparticle distances, to leading order in $(a/d)^3\ll1$, one can ignore the 
term $\propto(a/d)^6$ in the denominator of Eq.~\eqref{eq:chi_dimer},\footnote{Since in our model of interacting point dipoles the interparticle distance cannot be smaller than $d=3a$ \cite{park04_PRB}, we have $(a/d)^3\leqslant1/27\ll1$.} 
and write 
\begin{equation}
\label{eq:chi_approx}
\chi^{\sigma\sigma}\simeq\overline{\chi}+\Delta\chi^\sigma, \qquad
\Delta\chi^\sigma=-\eta_\sigma\frac{4\pi}{3}\chi_1\chi_2\left(\frac ad\right)^3.
\end{equation}
The sign of the interaction-induced correction term $\Delta\chi^\sigma$ thus depends on both the polarization $\sigma$ and the sign of the product $\chi_1\chi_2$. Due to the anisotropy of the dipole--dipole interaction, the absolute value of $\Delta\chi^\sigma$ is twice as much in the longitudinal direction as in the transverse one. In Fig.~\ref{fig:interaction_dimer} 
we represent by red arrows the magnetic configurations for the different cases that 
are slightly modified with respect to the previously discussed noninteracting situation (green arrows). 
For a transverse applied field (left column), the anti-alignment tendency of the dipolar coupling makes the magnetic response less paramagnetic when 
$\{\chi_1>0, \chi_2>0\}$ than in the noninteracting case [panel (a)], while diamagnetism is reinforced when
$\{\chi_1<0, \chi_2<0\}$ [panel (b)]. When $\{\chi_1>0, \chi_2<0\}$ [panel (c)], 
interactions increase the ZFS, but
the overall response depends on the sign of $\overline{\chi}$. 
For a longitudinally-applied magnetic field (right column in Fig.~\ref{fig:interaction_dimer}), the aligning tendency of the dipolar coupling
reinforces (weakens)
the paramagnetic (diamagnetic) response of the dimer when both nanoparticles are paramagnetic (diamagnetic) [panels (d) and (e), respectively]. 
In the case where the magnetic susceptibilities of the individual nanoparticles are opposite 
[panel (f)], the magnetic response of the interacting dimer depends on the relative strength of 
$\chi_1$ and $\chi_2$. Nevertheless, the negative $\Delta\chi^z$ indicates a tendency towards a more diamagnetic response.

Remarkably, when the 
full expression \eqref{eq:chi_dimer} of the ZFS of the interacting dimer is considered, and 
when both nanoparticles are either paramagnetic or diamagnetic (i.e., $\chi_1\chi_2>0$), there appears a divergence 
in $\chi^{\sigma\sigma}$ for the polarization-dependent critical value
$\chi_1\chi_2|_\mathrm{c}=({9}/{16\pi^2\eta_\sigma^2})(d/a)^6$. 
Such a divergence of the ZFS within our linear response approach
of Eq.~\eqref{eq:M_i_linear} marks the presence of a magnetic instability. 
In the case where the two nanoparticles have opposite magnetic responses ($\chi_1\chi_2<0$), no such instability is present.
Assuming that both nanoparticles have the same ZFS ($\chi_1=\chi_2=\chi$), we obtain the critical susceptibility 
\begin{equation}
\label{eq:chi_c}
\chi_\mathrm{c}^\sigma=\frac{3}{4\pi|\eta_\sigma|}\left(\frac da\right)^3, 
\end{equation}
which crucially depends on the interparticle separation distance (scaled with the nanoparticle radius) as $(d/a)^3$.
Considering an interparticle distance $d=3a$, we have
$\chi_\mathrm{c}^z\simeq3.22$ ($\chi_\mathrm{c}^{x,y}\simeq6.45$) for the longitudinal (transverse) direction. 
Such very high values are certainly not attainable in bulk materials, but we will later discuss if the large orbital magnetic response
of finite-size nanoparticles \cite{gomez18_PRB} can give enough scope to observe such a magnetic instability.

The above-mentioned divergence of the dimer ZFS \eqref{eq:chi_dimer}, 
signaling the presence of a magnetic instability,
results in an infinite magnetization of the dimer
at finite external field, which is obviously unphysical.
Such a behavior, following from  
the linear response assumption of Eq.~\eqref{eq:M_i_linear}, is not valid for large effective magnetic fields $H_i$. 
In the following, we shall introduce a model (before tackling the more
realistic microscopic description of Ref.~\cite{gomez18_PRB} in Sec.~\ref{sec:micro}) which yields a finite value of the total magnetization of the nanoparticle dimer, giving much insight into the physics at play.

\section{Saturating model}
\label{sec:toy}

In order to remedy for the above-mentioned unphysical behavior when both nanoparticles have 
either a paramagnetic or diamagnetic susceptibility larger (in absolute value) than $\chi_\mathrm{c}^\sigma$, we now go 
beyond the linear response assumption of Eq.~\eqref{eq:M_i_linear}, 
incorporating the fact that the nanoparticle magnetization typically saturates beyond a 
sufficiently large field (see Fig.~\ref{fig:M_1NP} describing microscopic calculations).
We then assume that the magnetic moment $\mathcal{M}_i$ of each nanoparticle saturates 
for large (effective) magnetic field to some finite value $\mathcal{M}_0$, which for simplicity is taken to be the same for both nanoparticles.
We then adopt for the functional form of $\boldsymbol{\mathcal{M}}_i$ in Eq.\ \eqref{eq:M_i} the expression
\begin{equation}
\label{eq:M_i_tanh}
\boldsymbol{\mathcal{M}}_i=\mathcal{M}_0\tanh{\left(\frac{\mathcal{V}\chi_iH_i}{\mathcal{M}_0}\right)}\hat{\mathbf{H}}_i, 
\end{equation}
which recovers the linear behavior of Eq.~\eqref{eq:M_i_linear} when ${\mathcal{V}\chi_iH_i}/{\mathcal{M}_0}\ll1$.
The set of self-consistent transcendental equations \eqref{eq:H_i} and \eqref{eq:M_i_tanh} can then be solved numerically using an 
iterative method. 

\begin{figure}[tb]
\begin{center}
\includegraphics[width=.9\linewidth]{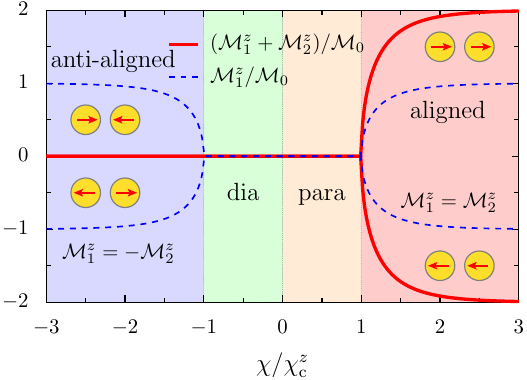}
\caption{\label{fig:magnetization_dimer}%
Red solid lines: Total magnetic moment $2m^z=\mathcal{M}_1^z+\mathcal{M}_2^z$ of the nanoparticle dimer at zero applied field ($H=0$)
in the longitudinal direction
obtained within the model of Eq.~\eqref{eq:M_i_tanh}
for an arbitrary interparticle spacing,
as a function of the zero-field susceptibility of the individual nanoparticles $\chi$ (assumed to be the same for both of them) in units of the critical susceptibility 
$\chi_\mathrm{c}^z$ from Eq.~\eqref{eq:chi_c}. The dashed blue lines show the magnetic moment of one individual nanoparticle.
The colored background indicates the different magnetic phases, i.e., for increasing values of $\chi$: anti-aligned (in blue), diamagnetic (in green), paramagnetic (in salmon), and aligned (in red), and corresponds to the color code in the phase diagram of Fig.~\ref{fig:phase_diagram}.}
\end{center}
\end{figure}

Assuming first that $\chi_1=\chi_2=\chi$ and a vanishing external field ($H=0$), 
we obtain the results of Fig.~\ref{fig:magnetization_dimer} for the $z$-component of the total magnetic moment $2m^z=\mathcal{M}_1^z+\mathcal{M}_2^z$ of the dimer in the longitudinal direction (red solid line) as a 
function of $\chi$ [scaled by the critical susceptiblity $\chi_\mathrm{c}^z$ from Eq.~\eqref{eq:chi_c} found within the linear response approach]. 
Notably, the only dependence of the results of Fig.~\ref{fig:magnetization_dimer} on the crucial parameter $d/a$ is through the critical susceptibility $\chi_\mathrm{c}^z$, and the zero-field total magnetic moment found only depends on the ratio $\chi/\chi_\mathrm{c}^z$. 
This is due to the fact that taking $H=0$ in the definition $\chi=\frac{1}{\mathcal{V}}\left.\frac{\partial\mathcal{M}}{\partial H}\right|_{H=0}$ of the ZFS puts us in the linear regime as long as there is no spontaneous magnetization.
Remarkably, when the two nanoparticles are paramagnetic ($\chi>0$), a finite magnetic moment (at vanishing external field) develops along the longitudinal direction, when $\chi$ is above the critical susceptibility $\chi_\mathrm{c}^z$. The magnetic moment eventually saturates to $m^z=\pm\mathcal{M}_0$ for large $\chi$. Of course, there is no symmetry breaking between the positive and negative magnetization states, since the latter are degenerate. While in large systems such degenerate states could be metastable and lead to ferromagnetic behavior, in our mesoscopic nanoparticle dimer the lifetimes of these states are not expected to be long enough for the emergence of a ferromagnetic hysteresis. Nevertheless, a magnetic supermoment with parallel magnetization of the two nanoparticles appears. In the presence of thermal fluctuations the resulting behavior is then superparamagnetic due to the large magnetic moment that can be oriented when an  external field breaks the degeneracy of the two magnetically-ordered states. 
Such an orbital-induced superparamagnetic behavior must not be confused with the one encoutered in single-domain nanoparticles made of a ferromagnetic material, where the magnetic anisotropy energy is too weak so as to ensure a permanent magnetization, even below the critical temperature~\cite{neel}.

As can be seen from the results displayed in Fig.~\ref{fig:magnetization_dimer}, when both nanoparticles are diamagnetic, there is no net magnetization for all negative values of $\chi$. However, the inspection of the magnetic response of each individual nanoparticle (dashed blue lines in Fig.~\ref{fig:magnetization_dimer}) shows that an anti-aligned order develops when $\chi<-\chi_\mathrm{c}^z$. 
For $|\chi|<\chi_\mathrm{c}^z$, 
there is no finite magnetization at zero field and 
the magnetic response of the dimer follows that of the individual nanoparticles: it is paramagnetic (diamagnetic) for 
$0<\chi<\chi_\mathrm{c}^z$ ($-\chi_\mathrm{c}^z<\chi<0$), as expected from the discussion in Sec.~\ref{sec:linear} [see in particular Fig.~\ref{fig:interaction_dimer}, panels (d) and (e), respectively].

From the result \eqref{eq:chi_c}, one could expect the appearance of a magnetic moment in the transverse direction (i.e., with $m^{x,y}\neq0$) at
susceptibilities $\chi$ above (in absolute value) $\chi_\mathrm{c}^{x,y}=2\chi_\mathrm{c}^z$. However, at such large values of $\chi$, both magnetic moments are already at saturation in the $z$-direction, so that the model of Eq.~\eqref{eq:M_i_tanh} yields $m^{x,y}\simeq0$.

Equipped with the above information, the self-consistent equations \eqref{eq:H_i} and \eqref{eq:M_i_tanh} reduce to 
\begin{equation}
\label{eq:Ising}
\frac{\mathcal{M}_i^z}{\mathcal{M}_0}=\tanh{\left(\frac{\chi}{\chi_\mathrm{c}^z} \frac{\mathcal{M}_j^z}{\mathcal{M}_0}\right)}, \qquad (i\neq j), 
\end{equation}
where symmetry dictates that $|\mathcal{M}_i^z|=|\mathcal{M}_j^z|$. When $\mathcal{M}_i^z=\mathcal{M}_j^z$, 
the graphical solution of Eq.~\eqref{eq:Ising} leads to
a magnetic order with parallel magnetic moments for $\chi>\chi_\mathrm{c}^z$, with $\mathcal{M}_i^z=m^z\simeq\pm\mathcal{M}_0\sqrt{3(\chi/\chi_\mathrm{c}^z-1)}$ for $\chi\rightarrow{\chi_\mathrm{c}^z}^+$ (see Fig.~\ref{fig:magnetization_dimer}).  
When $\mathcal{M}_i^z=-\mathcal{M}_j^z$, an anti-aligned magnetic order (for which $m^z=0$) develops for $\chi<-\chi_\mathrm{c}^z$ with 
$\mathcal{M}_i^z\simeq\pm\mathcal{M}_0\sqrt{-3(\chi/\chi_\mathrm{c}^z+1)}$ for $\chi\rightarrow{-\chi_\mathrm{c}^z}^-$.  

\begin{figure}[tb]
\begin{center}
\includegraphics[width=\linewidth]{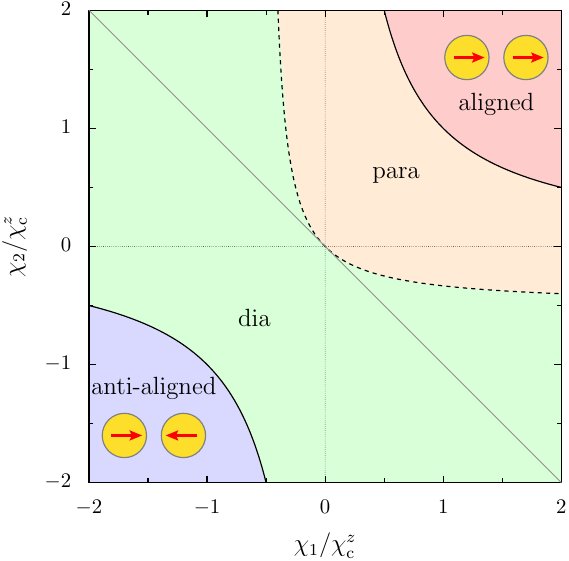}
\caption{\label{fig:phase_diagram}%
Phase diagram for the magnetic order in a dimer of nanoparticles in the longitudinal direction at vanishing external field ($H=0$) obtained 
numerically by solving Eqs.~\eqref{eq:H_i} and \eqref{eq:M_i_tanh}. 
The diamagnetic (in green) and paramagnetic regions (in salmon) are separated by the dashed line when the interparticle interaction is considered and by the thin gray line $\chi_2=-\chi_1$ when interactions are neglected. 
The blue and red regions correspond to the phases with anti-aligned and aligned orders, respectively.}
\end{center}
\end{figure}

\begin{figure*}[tb]
\begin{center}
\includegraphics[width=\linewidth]{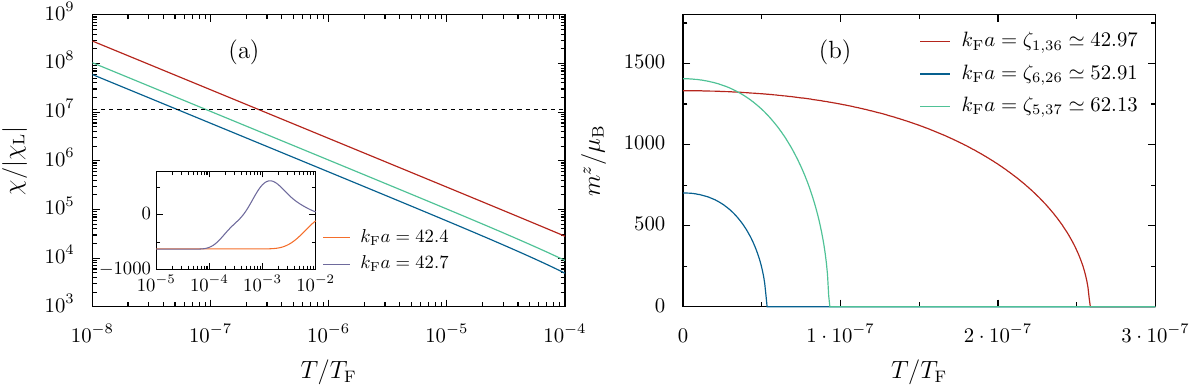}
\caption{\label{fig:dimer_real}%
(a) Main panel (paramagnetic behavior): Zero-field susceptibility of a single nanoparticle as a function of temperature for 
$k_\mathrm{F}a=\zeta_{1,36}\simeq42.97$ (red line), 
$k_\mathrm{F}a=\zeta_{6,26}\simeq52.91$ (blue line), 
and $k_\mathrm{F}a=\zeta_{5,37}\simeq62.13$ (green line).
The dashed black line corresponds to the critical susceptibility $\chi_\mathrm{c}^z$ [cf.\ Eq.~\eqref{eq:chi_c}] for an interparticle 
distance $d=3a$.
Inset (diamagnetic behavior at low temperature): Same as the main panel for $k_\mathrm{F}a=42.4$ (orange line) and $k_\mathrm{F}a=42.7$ (violet--gray line). 
(b) $z$-component of the total magnetic moment per particle 
$m^z$ at zero applied field ($H=0$) 
as a function of temperature for three values of $k_\mathrm{F}a$ for 
which both nanoparticles of the dimer have a paramagnetic response.}
\end{center}
\end{figure*}

Let us now consider the more general case in which the ZFSs of the individual nanoparticles are different, i.e., $\chi_1\neq\chi_2$. By solving numerically 
the set of self-consistent equations \eqref{eq:H_i} and \eqref{eq:M_i_tanh} for $H=0$, we find the phase diagram of Fig.~\ref{fig:phase_diagram} which
displays the magnetic phase in the longitudinal direction\footnote{In our iterative algorithm, when the magnetization of the dimer converges to zero, we subsequently apply a very weak external magnetic field so as 
to determine whether the magnetic response is para- or diamagnetic.}
as a function of $\chi_1$ and $\chi_2$.
The behavior along the diagonal $\chi_1=\chi_2$ is described in Fig.~\ref{fig:magnetization_dimer}.
The solid black lines delimitate 
the regions of the $\{\chi_1,\chi_2\}$ plane where a magnetic order 
develops (aligned moments for $\{\chi_1>0,\chi_2>0\}$ and anti-aligned for $\{\chi_1<0,\chi_2<0\}$). The functional form of such solid lines 
can be inferred from the linear model discussed in Sec.~\ref{sec:linear} and is given by $\chi_1\chi_2=(\chi_\mathrm{c}^z)^2$. 
The dashed line in Fig.~\ref{fig:phase_diagram} marks the separation between the paramagnetic and diamagnetic responses of the dimer. Here also, the functional form of the separation is determined by the linear model, and corresponds to $\chi^{zz}=0$ [cf.\ Eq.~\eqref{eq:chi_dimer}]. 
Interestingly, when $\chi_1$ and $\chi_2$ have opposite signs, the system tends to be more likely diamagnetic than paramagnetic. This can also be understood in terms of the linear response model, where in this case the interaction-induced term $\Delta\chi^z$ in Eq.~\eqref{eq:chi_approx} is negative [cf.\ Fig.~\ref{fig:interaction_dimer}(f)].

In the following, we will consider a more realistic, microscopic model \cite{gomez18_PRB} of the orbital magnetic response of a nanoparticle subject to an (effective) external field, and demonstrate that the transition to a magnetically-ordered aligned phase might be experimentally achievable, while the anti-aligned one remains elusive.

\section{Microscopic description of the electron dynamics}
\label{sec:micro}

\begin{figure}[tb]
\begin{center}
\includegraphics[width=\linewidth]{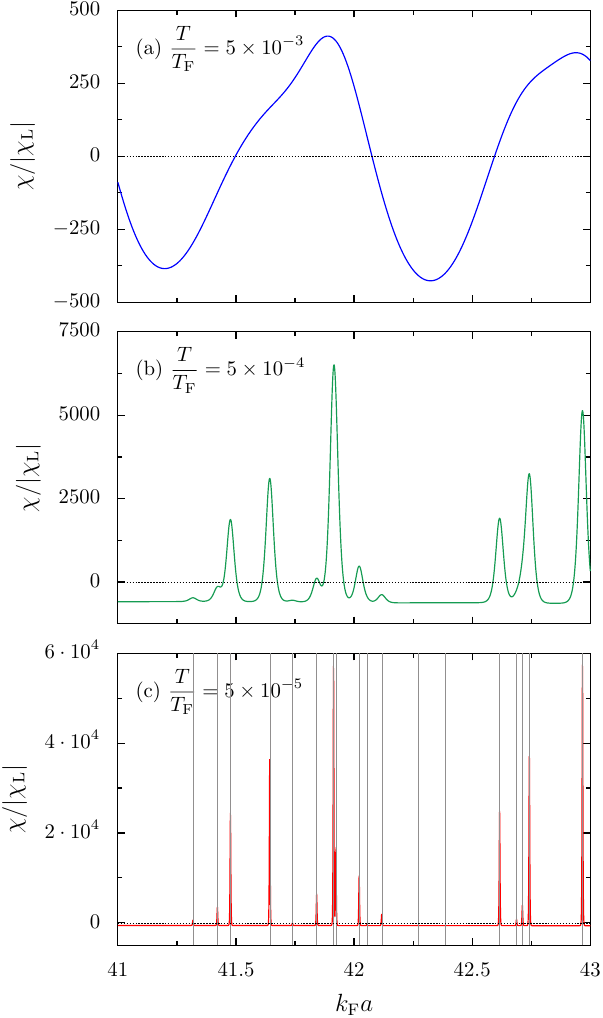}
\caption{\label{fig:ZFS_1NP}%
Colored solid lines: Zero-field susceptibility as a function of the nanoparticle radius $a$ (scaled with the Fermi wave vector $k_\mathrm{F}$) 
 for (a) $T/T_\mathrm{F}=5\times10^{-3}$ (corresponding to room temperature for gold), 
(b) $T/T_\mathrm{F}=5\times10^{-4}$, and (c) $T/T_\mathrm{F}=5\times10^{-5}$. The thin dashed black lines correspond to $\chi=0$. In panel (c), 
the vertical thin gray lines indicate the zeroes $\zeta_{nl}$ of the spherical Bessel functions [$j_l(k_\mathrm{F}a)=0$] for which one has a paramagnetic peak of the susceptibility.
}
\end{center}
\end{figure}

The saturating model discussed in the previous section imposes an arbitrary magnetic moment $\mathcal{M}_0$ at saturation, as well 
as the form \eqref{eq:M_i_tanh} of the individual nanoparticle magnetization. In order to gauge
the relevance of the above-discussed magnetic instabilities, it is important to relax the previous assumptions and consider the magnetic moments arising from the conduction electron orbital motion 
within each nanoparticle. In Appendix \ref{app:PRB18} we present the ZFS and the finite-field magnetization of individual nanoparticles, focusing on their temperature and size dependence. 
As illustrated in Fig.~\ref{fig:ZFS_1NP}(a), upon relatively modest changes of the particle radius (of the order of $k_\mathrm{F}^{-1}\sim\unit[1]{\AA}$), the ZFS at room temperature oscillates between paramagnetic and diamagnetic values which are typically much larger than $|\chi_\mathrm{L}|$.
For lower temperatures [Figs.~\ref{fig:ZFS_1NP}(b) and \ref{fig:ZFS_1NP}(c)], the paramagnetic component of the ZFS is signaled by large peaks associated with the highly 
degenerate $H=0$ spectrum of a spherical nanoparticle, which stick out from a 
smooth diamagnetic background.

The maximum values of $|\chi|$ obtained at room temperature are considerably smaller than the critical susceptibility \eqref{eq:chi_c} leading 
to the emergence of magnetic order. The higher susceptibilities obtained for lower temperatures advocate for a systematic study of the conditions under
which the critical susceptibility can be reached. Towards this goal we show 
in the main figure of Fig.~\ref{fig:dimer_real}(a) the evolution of the ZFS (cf.\ Eq.~(A5) in Ref.~\cite{gomez18_PRB}) with temperature for a few values of $k_\mathrm{F}a$ 
(some of them in the range shown in Fig.~\ref{fig:ZFS_1NP}): 
$k_\mathrm{F}a=\zeta_{1,36}\simeq42.97$ (red line), 
$k_\mathrm{F}a=\zeta_{6,26}\simeq52.91$ (blue line), 
and $k_\mathrm{F}a=\zeta_{5,37}\simeq62.13$ (green line) for which the ZFS is 
paramagnetic.\footnote{Here, $\zeta_{nl}$ is the $n$th zero of the spherical Bessel function of the first kind $j_l(z)$, and is related to the electronic eigenenergies at vanishing magnetic field, which locate the position (as a function of $k_\mathrm{F}a$) of the paramagnetic peaks of the ZFS, see Appendix \ref{app:PRB18} for further details.}
The inset in Fig.~\ref{fig:dimer_real}(a) presents the case of $k_\mathrm{F}a=42.4$ for which the ZFS is diamagnetic in the temperature interval considered (orange line), and that of $k_\mathrm{F}a=42.7$ 
for which the ZFS evolves from diamagnetic to paramagnetic as temperature increases 
(violet--gray line).
As can be seen from the main panel of Fig.~\ref{fig:dimer_real}(a), 
the paramagnetic ZFS dramatically increases with decreasing temperature, following a Curie-type law 
$\chi= \mathcal{C}(a)/T$, with $\mathcal{C}(a)$ a temperature-independent prefactor which depends on 
$a$.
In the diamagnetic case [inset of Fig.~\ref{fig:dimer_real}(a)], the temperature dependence of the ZFS is much less pronounced, and the attained values of $\chi\simeq -(2/5)(k_\mathrm{F}a)^2|\chi_\mathrm{L}|$  in the leading order of $k_\mathrm{F}a\gg1$ at low temperature \cite{gomez21_preprint}
remain much smaller (in absolute value) than in the paramagnetic case. 
Interestingly, for certain sizes (see the violet--gray solid line in the figure), 
the magnetic behavior can turn from paramagnetic at high temperatures to diamagnetic at lower temperatures, as can be inferred from the displayed results for the ZFS in Fig.~\ref{fig:ZFS_1NP}.

The previously discussed values of the ZFS in the diamagnetic case at low temperature are much smaller (in absolute value) than the required critical susceptibility $-\chi_\mathrm{c}^z$ needed 
to observe an anti-aligned magnetic order in the dimer. However, in the paramagnetic case, at very low temperatures 
(i.e., below $T/T_\mathrm{F}\sim10^{-7}$, corresponding to around $\unit[6]{mK}$ for gold), the ZFS can exceed $\chi_\mathrm{c}^z$ [which is shown by a dashed black line in Fig.~\ref{fig:dimer_real}(a) for an interparticle distance $d=3a$], such that a magnetically ordered phase with aligned moments may emerge. 

To check the above hypothesis, we solve the set of self-consistent equations \eqref{eq:M_i} and 
\eqref{eq:H_i} using 
the microscopically calculated individual magnetic moment from Eq.~(A4) in Ref.~\cite{gomez18_PRB}, and find the results of Fig.~\ref{fig:dimer_real}(b) for the TMMPP $m^z$ of the dimer in the longitudinal direction at zero applied field $(H=0)$, for nanoparticle sizes that correspond 
to the ZFS shown in panel (a) of the figure and for an interparticle distance $d=3a$.\footnote{We have checked numerically, using 
our iterative solution to the system of self-consistent equations \eqref{eq:M_i} and 
\eqref{eq:H_i} discussed above (together with Eq.~(A4) in Ref.~\cite{gomez18_PRB}), that the magnetization at vanishing applied field in the transverse direction 
is always zero. 
The results displayed in Fig.~\ref{fig:dimer_real}(b) are then obtained imposing $m^{x,y}=0$, 
which greatly facilitates the numerical calculations and improves its accuracy, as it boils down to a straightforward root-finding for the unknown variable $m^z$.} 
As can be seen from the figure, a spontaneous magnetization appears
below a critical temperature that corresponds to the crossing of the dashed line representing $\chi_\mathrm{c}^z$
for $d=3a$ and the displayed ZFSs, thus confirming the existence of a stable magnetically-ordered aligned phase. The values attained by $m^z$ at a temperature approaching the absolute zero can be as large as approximately a thousand times the Bohr magneton $\mu_\mathrm{B}$ for the considered nanoparticle sizes, and may thus be experimentally detectable.

\section{Conclusions}
\label{sec:ccl}

Several experimental works have reported over the last two decades
anomalous magnetic response of macroscopic assemblies of gold nanoparticles coated with organic ligands \cite{hori99_JPA, nakae00_PhysicaB, hori04_PRB, yamam04_PRL, guerr08, guerr08_Nanotechnology, barto12_PRL, agrac17_ACSOmega, cresp04_PRL, dutta07_APL, rhee13_PRL, cresp06_PRL, donni07_AdvMater, garit08_NL, venta09, donni10_SM, maitr11_CPC, grege12_CPC}.
Motivated in particular by the occurrence of the yet unexplained ferromagnetic behavior
 \cite{cresp04_PRL, cresp06_PRL, dutta07_APL, donni07_AdvMater, garit08_NL, guerr08_Nanotechnology, guerr08, venta09, donni10_SM, maitr11_CPC, agrac17_ACSOmega, grege12_CPC}, we have posed the question whether 
the combined effects of largely-enhanced orbital magnetism (due to strong quantum confinement \cite{gomez18_PRB}) and dipolar interactions can lead to ferromagnetic order. 
Attempting to answer this question, we have focused on 
the building block of any realistic sample, i.e., a dimer of spherical nanoparticles. 

Using firstly a linear-response approach, where the magnetic moment of each nanoparticle in the dimer is proportional to the local magnetic field (i.e., the applied one plus that generated by the other nanoparticle), and secondly a model in which each magnetic moment saturates at large fields, we have unveiled the existence of both, a magnetically-ordered aligned phase (when both particles present a large paramagnetic ZFS), and an anti-aligned magnetic order (when the two particles have a large diamagnetic ZFS). For weaker ZFSs, we have shown that interactions tend to favor a diamagnetic behavior. Thirdly, using the microscopic quantum description of Ref.~\cite{gomez18_PRB}, we have demonstrated that the transition to the aligned phase may be experimentally observable at cryogenic temperatures and for precise nanoparticle sizes (where the individual paramagnetic ZFSs are way above the Landau susceptibility), while the anti-aligned order seems unlikely to be reached. 

The mesoscopic dipoles with very large magnetic moments were treated with a purely 
energetic approach, neglecting thermal fluctuations. The validity of such an approach is 
supported by the heuristic model of Appendix \ref{app:dimer}, where the thermal equilibrium
of a dimer of interacting particles with variable magnetic moments results in a superparamagnetic behavior beyond a critical value of the interaction. 

While we limited our analytical and numerical calculations to center-to-center interparticle distances $d$ larger than $3a$ (where $a$ is the nanoparticle radius), where the static magnetic dipolar interactions are 
dominating and where we have shown that the critical susceptibility above which a magnetic moment at zero applied field appears and scales as $(d/a)^3$, we expect 
that for $d<3a$, the critical value would further decrease. However, 
a reliable description in such a regime would require the inclusion of 
multipolar interactions, which is beyond the scope of the present work. 

Interestingly, when one considers a linear chain of identical paramagnetic nanoparticles, the critical ZFS above which one may observe magnetic order and a spontaneous magnetization decreases with increasing chain length, and saturates to a plateau which is about a factor of two smaller than in the case of a dimer (see Appendix \ref{app:chain}). This tendency favoring the alignment of the mesoscopic magnetic moments in the longitudinal configuration follows from the increase of the number of neighbors and may facilitate the experimental observation of the spontaneous orbital magnetization. 

It is worth mentioning that we do not expect our predictions to be restricted to the spherical nanoparticle shape, as many other three-dimensional classically-integrable geometries such as cuboids \cite{ruite91_PRL, leuwe93_PhD}, half-spheres \cite{fraue98_PRB, gomez21_preprint} or cylinders \cite{gomez_unpublished} present a large orbital magnetic susceptibility, similar to the case of two-dimensional semiconducting heterostructures \cite{richt96_PhysRep}. It should however be noticed that classically-chaotic ballistic systems present a somewhat weaker orbital response \cite{richt98_EPL}, so that a regular nanoparticle geometry is to be favored for the observation of the effects we studied here. 

The experimental detection of our theoretical predictions requires both very precise nanofabrication and detection techniques, as well as the use of cryogenic temperatures. 
On the one hand, there has been many recent advances in the elaboration of atomically precise ligand-protected Au clusters \cite{Bonacchi} (notably with a large paramagnetic response \cite{agrac17_ACSOmega}). On the other hand, the recent proposal of Ref.~\cite{roda21_PRB} for detecting the magnetic response of single nanoparticles using superconducting quantum interference devices may be extended to the dimers and chains considered in this work.

Due to the above-mentioned stringent experimental requirements for observing in nanoparticle dimers the appearance at very low temperature of an interaction-induced spontaneous magnetization and a reinforced diamagnetic phase, it is quite unlikely
that the present mechanism can explain the experiments of Refs.~\cite{hori99_JPA, nakae00_PhysicaB, hori04_PRB, yamam04_PRL, guerr08, guerr08_Nanotechnology, barto12_PRL, agrac17_ACSOmega, cresp04_PRL, dutta07_APL, rhee13_PRL, cresp06_PRL, donni07_AdvMater, garit08_NL, venta09, donni10_SM, maitr11_CPC, grege12_CPC} 
on three-dimensional assemblies of nanoparticles. Indeed, such experiments are performed with particles that present a relatively large size dispersion. Moreover, the samples are most likely disordered, which induces magnetic frustration. Finally, some of these experiments reported a ferromagnetic instability up to room temperature. The interpretation of the unusual magnetic behavior of gold nanoparticle assemblies thus remains an open problem.

\section*{Acknowledgements}
We thank Mauricio G\'omez Viloria and Yves Henry for helpful discussions. 
We acknowledge financial support from the French National Research Agency (ANR) through Grant No.\ ANR-14-CE26-0005 Q-MetaMat. 
This work of the Interdisciplinary Thematic Institute QMat, as part of the ITI 2021-2028 program of the University of Strasbourg, CNRS, and Inserm, was supported by IdEx Unistra (ANR 10 IDEX 0002), and by SFRI STRAT’US project (ANR 20 SFRI 0012) and EUR QMAT ANR-17-EURE-0024 under the framework of the French Investments for the Future Program.

\begin{appendix}
\section{Zero-field susceptibility and finite-field magnetic moment of an individual nanoparticle}
\label{app:PRB18}

Here we summarize the physical assumptions and reproduce some quantum-mechanical results used 
in Ref.\ \cite{gomez18_PRB} to obtain the orbital magnetic response of a single spherical metallic nanoparticle useful
for the discussion carried in the main text.

\begin{figure}[tb]
\begin{center}
\includegraphics[width=\linewidth]{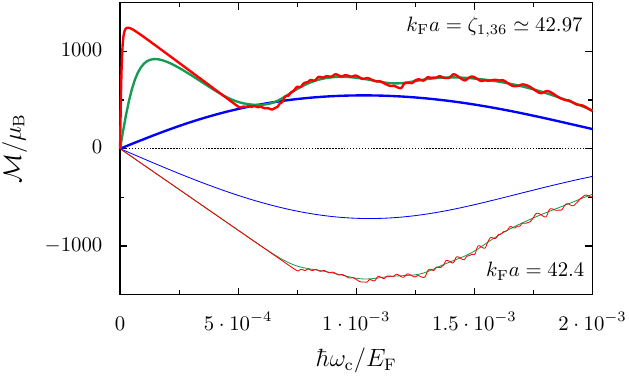}
\caption{\label{fig:M_1NP}%
Colored solid lines: Magnetic moment $\mathcal{M}$ as a function of the cyclotron energy $\hbar\omega_\mathrm{c}\sim H$ (scaled with the Fermi energy $E_\mathrm{F}$) for $T/T_\mathrm{F}=5\times10^{-3}$ (blue lines), 
$T/T_\mathrm{F}=5\times10^{-4}$ (green lines), and $T/T_\mathrm{F}=5\times10^{-5}$ (red lines). 
The thick (thin) lines correspond to $k_\mathrm{F}a=\zeta_{1,36}\simeq42.97$ ($k_\mathrm{F}a=42.4$) for which the 
zero-field susceptibility is paramagnetic (diamagnetic), see Figs.~\ref{fig:dimer_real} and \ref{fig:ZFS_1NP}.
}
\end{center}
\end{figure}

In a nutshell, the model of Ref.~\cite{gomez18_PRB} 
considers a spherical nanoparticle of radius $a$ where the ionic background is treated as a positively-charged jellium. The inclusion of 
electron-electron interactions at mean-field level leads to an effective self-consistent potential for the valence electrons 
that can be approximated by a hard-wall potential \cite{weick05_PRB}. The considered size of the nanoparticle is such that electronic correlations and disorder effects can 
be disregarded. The spin--orbit coupling, as well as other relativistic corrections are neglected since they do not participate significantly to the magnetic response 
\cite{gomez21_preprint}. The ligands surrounding the nanoparticles are ignored, since they do not seem to play a significant role on the experimental 
results \cite{nealo12_Nanoscale}.\footnote{In the model of Ref.~\cite{gomez18_PRB}, the slight difference between the applied field $H$ and the magnetic induction $B=H+4\pi M$, with $M=\mathcal{M}/\mathcal{V}$
the magnetization density within the nanoparticle, is not taken into account, as $4\pi M\ll H$, even when the magnetic moment reaches values as large as $\mathcal{M}\sim10^3\mu_\mathrm{B}$, as is the case, e.g., in the results displayed in Fig.~\ref{fig:M_1NP}.}$^,$\footnote{The number of electrons $N  \simeq 4(k_\mathrm{F}a)^3/9\pi$ 
in each particle being fixed, one should in principle work within the canonical ensemble
when determining the magnetic response of individual nanoparticles, as is the case in this paper. 
Working with the grand-canonical ensemble 
introduces relative 
errors of the order of $N^{-1/2}$, which are negligible for not too small nanoparticles. The
canonical corrections become important for an ensemble of nanoparticles, where the
average over
different sizes can lead to a vanishing grand-canonical result \cite{gomez18_PRB}, 
as is the case for persistent currents in mesoscopic rings \cite{bouch89_JP}.
In this work, dealing with a nanoparticle dimer, 
we are not concerned with these corrections.}

The model above described, when treating perturbatively the diamagnetic term of the corresponding Hamiltonian, results in the ZFS given by Eq.~(A5) in Ref.~\cite{gomez18_PRB}
and presented in Fig.~\ref{fig:ZFS_1NP}. We there show the ZFS as a function of the nanoparticle radius $a$ (scaled with the Fermi wave vector $k_\mathrm{F}$) 
for three different temperatures (scaled with the Fermi temperature $T_\mathrm{F}$): $T/T_\mathrm{F}=5\times10^{-3}$ which corresponds 
to the case of gold at room temperature [Fig.~\ref{fig:ZFS_1NP}(a)], 
$T/T_\mathrm{F}=5\times10^{-4}$ [Fig.~\ref{fig:ZFS_1NP}(b)], and $T/T_\mathrm{F}=5\times10^{-5}$ [Fig.~\ref{fig:ZFS_1NP}(c)]. 
At room temperature, one can see on Fig.~\ref{fig:ZFS_1NP}(a) that the ZFS oscillates as a function of the nanoparticle size between para- and diamagnetic values that are much larger (in absolute value) than the Landau susceptibility $\chi_\mathrm{L}$ given in Eq.~\eqref{eq:chi_L}. 
Since the ZFS of bulk gold is of the order of $10\,\chi_\mathrm{L}$, the quantum confinement of the electronic eigenstates has already a sizeable 
effect on the magnetic response of the nanoparticle at room temperature. 
When the temperature is lowered by one order of magnitude [Fig.~\ref{fig:ZFS_1NP}(b)], paramagnetic peaks on top of a diamagnetic background 
(of the order of $ -(2/5)(k_\mathrm{F}a)^2|\chi_\mathrm{L}|$ \cite{gomez21_preprint})
start to develop around certain values of $k_\mathrm{F}a$. For the displayed nanoparticle sizes, such peaks can attain values of the ZFS that are one order of magnitude larger than the typical values at room temperature. 
At even lower temperature [Fig.~\ref{fig:ZFS_1NP}(c)], 
these paramagnetic peaks are two orders of magnitude larger than the typical values of $\chi$ obtained at room temperature, 
and coincide with $k_\mathrm{F}a=\zeta_{nl}$. Here, $\zeta_{nl}$ corresponds to the $n$th zero of the spherical Bessel function of the first kind $j_l(z)$ [see the thin gray vertical lines in Fig.~\ref{fig:ZFS_1NP}(c)], and determines the zero-field spectrum characterized by the principal ($n$) and angular
momentum ($l$) quantum numbers \cite{gomez18_PRB}. 

In Fig.~\ref{fig:M_1NP} we show the magnetic moment $\mathcal{M}$ 
(cf.\ Eq.~(A4) in Ref.~\cite{gomez18_PRB}) as a function of the cyclotron energy 
$\hbar\omega_\mathrm{c}=\hbar e H/m_*c$ (scaled with the Fermi energy $E_\mathrm{F}$)\footnote{For gold nanoparticles, the maximal value of $\hbar\omega_\mathrm{c}/E_\mathrm{F}$ considered in Fig.~\ref{fig:M_1NP} 
corresponds to a very large magnetic field of $H=\unit[9.2\times10^5]{G}$.}
and for the three temperatures considered 
in Fig.~\ref{fig:ZFS_1NP}, displaying $\mathcal{M}$ for two values of $k_\mathrm{F}a$ considered in Fig.~\ref{fig:dimer_real} ($k_\mathrm{F}a=42.4$, thin solid lines, diamagnetic ZFS, and $k_\mathrm{F}a=\zeta_{1,36}\simeq42.97$, thick solid lines, paramagnetic ZFS). 
The linear response of $\mathcal{M}$ at weak magnetic field follows the above-discussed behavior of the ZFS
as a function of temperature in both the paramagnetic and diamagnetic cases (see Fig.~\ref{fig:dimer_real}). 
For larger applied fields, $\mathcal{M}$ can present a nonmonotonic behavior as a function of $H$, with 
typical values of several hundreds of the Bohr magneton~$\mu_\mathrm{B}$.

\section{Magnetic order in a model of two interacting paramagnetic nanoparticles} 
\label{app:dimer}

In this appendix we discuss a heuristic model of two interacting paramagnetic systems. 
We aim at approaching the situation of orbital magnetism in metallic nanoparticles where a large number of electrons contribute to the magnetic moment which is relatively insensitive to thermal fluctuations. We thus go beyond the usual description of paramagnetism where the magnetization results from the combined effect of an external field and thermal fluctuations on magnetic moments of fixed absolute value. Within our model, we find a transition to a large magnetic moment even in the absence of an external field,
in analogy with the behavior found and discussed in the main text.

\subsection{Magnetic response of a single nanoparticle with variable magnetic moment}
\label{app:dimer1}
We start by considering a heuristic model for a single nanoparticle with a magnetic moment 
$\boldsymbol{\mathcal{M}}$ that can vary in both, absolute value and orientation. In an external magnetic field $\mathbf{H}$, the magnetization-dependent contribution to the energy is assumed to have the form
\begin{equation}
\label{eq:E_1}
E_1(\boldsymbol{\mathcal{M}},\mathbf{H})=\gamma\boldsymbol{\mathcal{M}}^2-\boldsymbol{\mathcal{M}}\cdot\mathbf{H}.
\end{equation} 
The first term on the right-hand side of the above equation describes a preference for a low magnetic moment, whose strength is governed by the parameter $\gamma>0$, and the second term is the usual Zeeman potential energy. For a single nanoparticle it is straightforward to calculate the partition function as an integral over all values of $\boldsymbol{\mathcal{M}}$, expressed in spherical coordinates by the absolute value $\mathcal{M}=|\boldsymbol{\mathcal{M}}|$ and the associated polar and azimuthal angles, $\theta$ and $\phi$, respectively, as
\begin{align}
Z_1 &= \int_0^\infty \mathrm{d}\mathcal{M}\, \mathcal{M}^2
\int_0^\pi \mathrm{d}\theta\,  \sin{\theta}
\int_0^{2\pi}\mathrm{d}\phi\; 
\mathrm{e}^{-\beta E_1(\boldsymbol{\mathcal{M}},\mathbf{H})} 
\nonumber\\
&= \left(\frac{\pi}{\beta\gamma}\right)^{3/2}\exp\left(\frac{\beta H^2}{4\gamma}\right).
\end{align}  
The expectation value of the magnetic moment at thermal equilibrium along the direction $z$ of the magnetic field is then given by
\begin{equation}\label{eq:app1M1}
\left\langle \mathcal{M}^z \right\rangle =\frac{H}{2\gamma}.
\end{equation}
Remarkably, this expectation value and the corresponding magnetic susceptibility 
\begin{align}\label{eq:app1chi1}
\chi_1&= \frac{1}{\mathcal{V}}\frac{\partial\!\left\langle\mathcal{M}^z\right\rangle}{\partial H}
\nonumber\\
&=\frac{1}{2\mathcal{V}\gamma}
\end{align}
are independent of the temperature, and determined solely by the parameter $\gamma$. In such a model, the origin of the pa\-ra\-mag\-netic behavior is the magnetic-field induced absolute value of the magnetic moment rather than the usual orientation of a fixed modulus moment by the Zeeman term, in competition with thermal fluctuations in the case of a fixed modulus moment.\footnote{A temperature dependence will occur when the modulus of the magnetic moment is limited by an upper bound 
$\mathcal{M}_0$. Equation \eqref{eq:app1chi1} remains nevertheless a good approximation for the ZFS, provided $\sqrt{\beta\gamma}\mathcal{M}_0\gg 1$.} The fluctuations of the magnetic moment $\boldsymbol{\mathcal{M}}$ around its expectation value \eqref{eq:app1M1} are thus weak at low temperatures, even if the magnetization is small.

\subsection{Magnetic behavior of two interacting moments}
\label{app:dimer2}
We now consider the magnetic behavior of a dimer of nanoparticles placed at a distance $d$ along the $z$-axis (see Fig.~\ref{fig:sketch_dimer}), each of them described by the heuristic model of the previous Sec.~\ref{app:dimer1}, and coupled by the dipole--dipole interaction energy originating from their magnetic moments
\begin{equation}
V(\boldsymbol{\mathcal{M}}_1,\boldsymbol{\mathcal{M}}_2)=
\frac{\boldsymbol{\boldsymbol{\mathcal{M}}_1\cdot \mathcal{M}}_2-3\mathcal{M}_1^z\mathcal{M}_2^z}{d^3} .
\end{equation}
The magnetization-dependent part of the total energy of such a dimer, subject to an external magnetic field $\mathbf{H}$, can then be written in the form
\begin{align}
E_2(\boldsymbol{\mathcal{M}}_1,\boldsymbol{\mathcal{M}}_2,\mathbf{H}) =&\; 
E_1(\boldsymbol{\mathcal{M}}_1,\mathbf{H}) + E_1(\boldsymbol{\mathcal{M}}_2,\mathbf{H}) 
\nonumber\\
&+
V(\boldsymbol{\mathcal{M}}_1,\boldsymbol{\mathcal{M}}_2)
\end{align} 
as a function of the magnetic moments $\boldsymbol{\mathcal{M}}_1$ and $\boldsymbol{\mathcal{M}}_2$ of the two individual nanoparticles, and where $E_1$ is defined in Eq.~\eqref{eq:E_1}.

The partition function of the dimer model system can then be written as an integral over the absolute values $\mathcal{M}_i$ and the solid angles 
$\Omega_i = \{\theta_i,\phi_i\}$ of the magnetic moments ($i=1,2$) as 
\begin{align}
\label{eq:appz2}
Z_2 =&\; 
\int_0^{\mathcal{M}_0} \mathrm{d}\mathcal{M}_1\, \mathcal{M}_1^2
\int_0^{\mathcal{M}_0} \mathrm{d}\mathcal{M}_2\, \mathcal{M}_2^2
\nonumber\\
&\times
\int\mathrm{d}\Omega_1
\int\mathrm{d}\Omega_2\; 
\mathrm{e}^{-\beta E_2(\boldsymbol{\mathcal{M}}_1,\boldsymbol{\mathcal{M}}_2,\mathbf{H})} .
\end{align} 
As discussed in Sec.~\ref{sec:toy}, 
the existence of a saturation value of the magnetization is expected on physical grounds, and this is why in the above equation 
we introduced an upper limit $\mathcal{M}_0$ in the integrals over the magnitude of the magnetic moments.     
We assume that the external magnetic field is applied along the $z$-direction and introduce the reduced magnetic moments $\mu_i=\mathcal{M}_i/\mathcal{M}_0$. Then, an analytical evaluation is possible for two of the six integrals in Eq.~\eqref{eq:appz2}, and the partition function $Z_2$ can be expressed as 
\begin{align}\label{eq:appz2bis}
Z_2(h,K,C) =&\; 4\pi^2\mathcal{M}_0^6
\int_0^1 \mathrm{d}\mu_1\, \mu_1^2\, \mathrm{e}^{-C \mu_1^2}
\int_0^1 \mathrm{d}\mu_2\, \mu_2^2\, \mathrm{e}^{-C \mu_2^2}
\nonumber\\
&\times
\int_{-1}^{+1}\mathrm{d}u_1\, \mathrm{e}^{h \mu_1 u_1}
\nonumber\\
&\times
\int_{-1}^{+1}\mathrm{d}u_2\, \mathrm{e}^{h \mu_2 u_2} 
\mathrm{e}^{2K \mu_1 \mu_2 u_1 u_2}
\nonumber
\\
&\times   
I_0\!\left( K \mu_1 \mu_2 \sqrt{(1-u_1^2)(1-u_2^2)}\right)
\end{align}
in terms of four integrals that can be calculated numerically. 
We have defined the temperature-dependent dimensionless parameters 
$h=\beta \mathcal{M}_0 H$, $K=\beta \mathcal{M}_0^2/d^3$, and $C= \beta \gamma\mathcal{M}_0^2$, while $I_0(z)$ denotes the modified Bessel function of the first kind and of order $0$.

A similar integral with an additional factor of $\mu_1$ in the integrand, normalized with the partition function $Z_2$, allows us to compute the expectation value $\langle \mathcal{M}_1 \rangle$, related to the magnetic moment of the first nanoparticle in the dimer. In the absence of an external field, $H=0$, such an expectation value vanishes since any positive contribution to the integral is compensated by a symmetrical negative one. 

\begin{figure}
\begin{center}
\includegraphics[width=\linewidth]{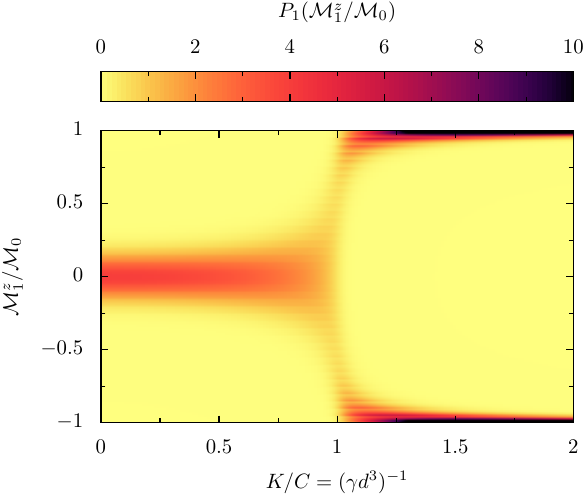}
\caption{\label{fig:appP1z}%
Probability density in thermal equilibrium of the $z$-component of the magnetic moment of the first nanoparticle of the dimer as a function of the interaction strength parameter $K$, scaled with $C$. The distributions, that are shown in vertical cuts in colorscale, have been evaluated at low temperature such that $C=50$.}
\end{center}
\end{figure}

The corresponding numerically-evaluated probability density in thermal equilibrium $P_1$ 
of having a magnetic moment $\mathcal{M}_1^z$
is shown in Fig.~\ref{fig:appP1z} for $C = 50$ as a function of the interaction strength parameter $K$, scaled with the value of $C$. While the probability density remains symmetric around zero magnetic moment in the $z$-direction, a striking change in behavior occurs when the interaction strength increases beyond the value at which $\gamma d^3=1$. While the probability density of the magnetization of the first nanoparticle has large values around  zero magnetization below the interaction threshold, such a quantity exhibits two peaks for saturated magnetic moments in positive and negative $z$-direction above the threshold.
Therefore, Fig.~\ref{fig:appP1z} presents, at the qualitative level, a similarity with 
the $\chi>0$ sector of Fig.~\ref{fig:magnetization_dimer} discussed in Sec.~\ref{sec:toy}.
Remembering that within our model the ZFS of a single particle \eqref{eq:app1chi1} is given by $\chi_1=(2\mathcal{V}\gamma)^{-1}$, the threshold corresponds to a value of the susceptibility of $\chi_{1,\mathrm{c}} =d^3/2 \mathcal{V}$, and coincides with the value of Eq.\ \eqref{eq:chi_c} found in Sec.\ \ref{sec:linear} for the critical susceptibility $\chi_\mathrm{c}^z$ in the $z$-direction. Above the threshold, we observe the emergence of an interaction-induced magnetic moment that corresponds to aligned and saturated moments of the two dipoles. If one adds an external magnetic field, this large total moment will then result in a superparamagnetic behavior.

Our heuristic model, allowing for a variable magnetic moment in each nanoparticle, 
demonstrates the reduced role of thermal fluctuations at equilibrium, 
with the corresponding ordering of mesoscopic dimers beyond a critical value 
of the interparticle interaction. Such a behavior should be contrasted with the case of a dimer
constituted by two interacting magnetic moments of fixed magnitude, where the 
magnetic response is always paramagnetic and increasing the interparticle 
interaction continuously increases the magnetic stiffness of the system.

\section{Magnetic order in the longitudinal direction of a chain of paramagnetic nanoparticles}
\label{app:chain}

\begin{figure}[tb]
\begin{center}
\includegraphics[width=\linewidth]{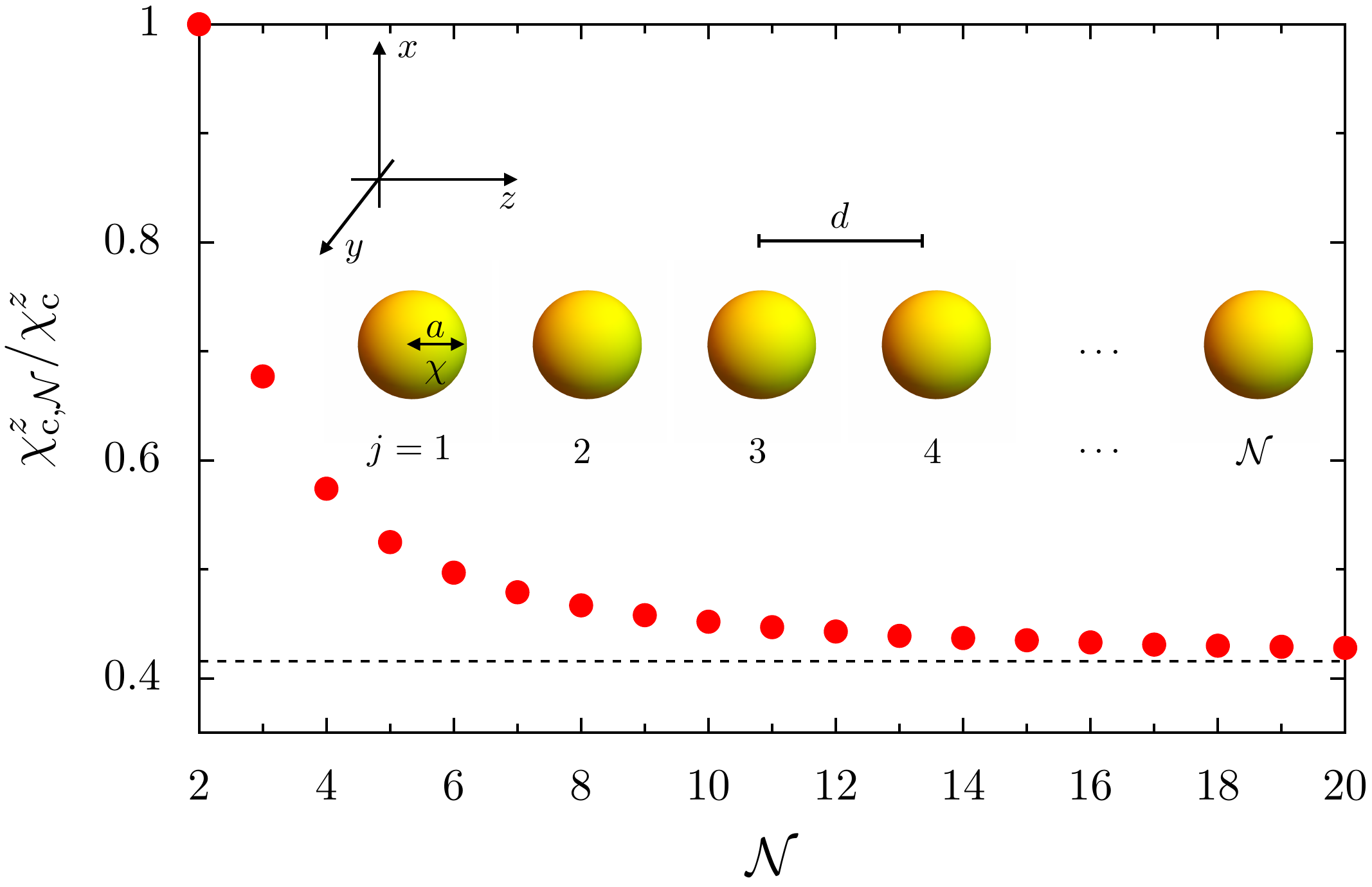}
\caption{\label{fig:chain_chi}%
Red dots: Critical susceptibility $\chi_{\mathrm{c}, \mathcal{N}}^z$ [scaled by the dimer result $\chi_{\mathrm{c}}^z$, see Eq.~\eqref{eq:chi_c}] as a function of the number $\mathcal{N}$ of nanoparticles in the chain above which aligned magnetic moments appear in the longitudinal direction.
Black dashed line: Asymptotic value of the critical zero field susceptibility for large $\mathcal{N}$, 
$\chi^z_{\mathrm{c}, \infty}=\chi^z_\mathrm{c}/2\zeta(3)\simeq 0.42\chi^z_\mathrm{c}$, where $\zeta(3)\simeq1.2$ is Ap\'ery's constant.
Inset: Sketch of a chain of $\cal N$ identical spherical metallic nanoparticles with radius $a$ and zero-field paramagnetic susceptibility $\chi>0$, separated by a center-to-center distance~$d$.}
\end{center}
\end{figure}

The finite magnetic moment at vanishing applied magnetic field found in the longitudinal configuration using the microscropic model of Sec.~\ref{sec:micro} [see in particular Fig.~\ref{fig:dimer_real}(b)]
in the case of a nanoparticle dimer with individual paramagnetic ZFSs calls for a more systematic study of such an instability as a function of system size. Therefore, we search for the conditions favoring the ordered phase with finite magnetic moments. 
In this appendix, we thus extend the results obtained for a mesoscopic dimer 
to the case of a linear chain comprising $\mathcal{N}$ identical metallic nanoparticles aligned along the $z$-axis, with a radius $a$, and separated by a center-to-center distance $d$, as sketched in the inset of Fig.~\ref{fig:chain_chi}. 
The nanoparticles are assumed to have the same size and thus the same 
orbital ZFS (see Appendix \ref{app:PRB18}), which we take it to be paramagnetic ($\chi>0$).

Within the linear-response approach of Sec.~\ref{sec:linear}, the set of self-consistent equations \eqref{eq:H_i} 
and \eqref{eq:M_i_linear} can be straightforwardly generalized to the case of $\mathcal{N}$ interacting
nanoparticles. This leads for vanishing external magnetic field ($H=0$) and for the longitudinal configuration ($\sigma=z$) to the system of linear equations 
\begin{equation}
\label{eq:M_N}
\mathcal{M}_i^z-\frac{\chi}{\chi_\mathrm{c}^z}
\sum_{\substack{j=1\\(i\neq j)}}^\mathcal{N}\frac{\mathcal{M}_j^z}{|i-j|^3}=0,\qquad (i=1,\ldots, \mathcal{N}),
\end{equation}
where $\mathcal{M}_i^z$ is the $z$-component of the magnetic moment 
of the $i$th nanoparticle in the chain.
The set of equations \eqref{eq:M_N} defines an $\mathcal{N}\times\mathcal{N}$ matrix and has nonvanishing solutions $\mathcal{M}_i^z\neq0$ only if its determinant is zero, thus determining the critical susceptibility $\chi_{\mathrm{c}, \mathcal{N}}^z$ above which aligned magnetic moments appear in the chain of $\mathcal{N}$ nanoparticles. 

In Fig.~\ref{fig:chain_chi} we display our results for $\chi_{\mathrm{c}, \mathcal{N}}^z$ as a function of $\mathcal{N}$ by red dots.
As can be seen from the figure, $\chi^z_{\mathrm{c}, \mathcal{N}}$ decreases as a function of $\mathcal{N}$ to reach a plateau for which it is about a factor of two smaller than the critical ZFS of the dimer $\chi_{\mathrm{c}}^z$ [cf.\ Eq.~\eqref{eq:chi_c}]. 
While the plateau is approached for modest values of $\mathcal{N}$, an analytic result can be obtained in the limit $N \gg 1$. Performing the Fourier transform of Eq.~\eqref{eq:M_N}, it is straightforward to demonstrate that the saturating value of $\chi^z_{\mathrm{c}, \mathcal{N}}$ is given by $\chi^z_{\mathrm{c}, \infty}=\chi^z_\mathrm{c}/2\zeta(3)\simeq 0.42\chi^z_\mathrm{c}$, where $\zeta(3)$ is Ap\'ery's constant. Such an asymptotic value of the critical ZFS is shown by a dashed line in Fig.~\ref{fig:chain_chi}.
The tendency favoring the appearance and the alignment of the magnetic moments in the longitudinal configuration is a consequence of the increase of the number of other particles with which one nanoparticle within the chain interacts through the long-range magnetic dipolar coupling. 

\begin{figure*}[tbh]
\begin{center}
\includegraphics[width=13truecm]{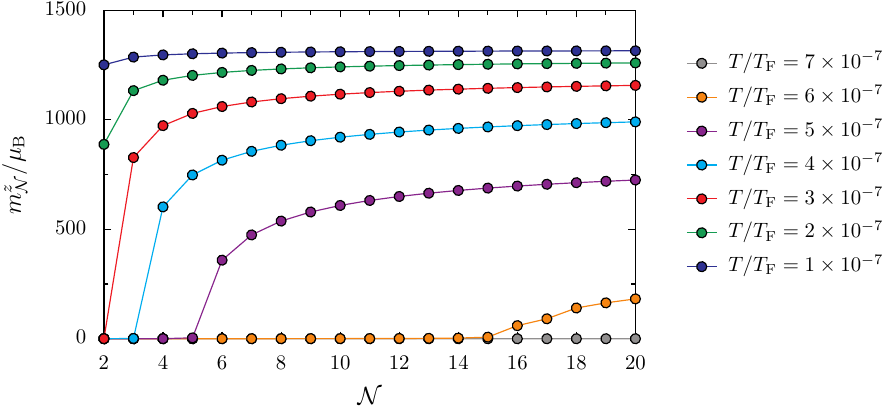}
\caption{\label{fig:chain_M}%
Colored dots: $z$-component of the zero-field total magnetic moment per particle $m^z_\mathcal{N}=\sum_{j=1}^\mathcal{N}\mathcal{M}_j^z/\mathcal{N}$ (scaled with the Bohr magneton $\mu_\mathrm{B}$) as a function of the chain length $\mathcal{N}$ and for increasing temperature $T$ (in units of the Fermi temperature $T_\mathrm{F}$) obtained using the microscopic quantum-mechanical model of Sec.~\ref{sec:micro} and Appendix \ref{app:PRB18}. The parameters are 
$k_\mathrm{F}a=\zeta_{1,36}\simeq 42.97$ and $d=3a$. In the figure, the solid lines are guides to the eyes.
}
\end{center}
\end{figure*}

To gauge the relevance of the above considerations deduced from the linear-response approach, we now adopt the microscopic quantum-mechanical model of Sec.~\ref{sec:micro} (see Appendix \ref{app:PRB18} for details) and show in Fig.~\ref{fig:chain_M} the TMMPP $m^z_\mathcal{N}=\sum_{j=1}^\mathcal{N}\mathcal{M}_j^z/\mathcal{N}$ resulting from such a model. 
As can be seen from Fig.~\ref{fig:chain_M}, the spontaneous TMMPP (i.e., for $H=0$) increases 
as a function of $\mathcal{N}$. Moreover, it can be concluded from the results of Fig.~\ref{fig:chain_M} that, as expected, lower temperatures favor the appearance of aligned magnetic moments within the chain.

As discussed at the end of Appendix \ref{app:dimer}, we stress here the difference between the above considered case of a chain of mesoscopic dipoles with that of microscopic moments, where thermal fluctuations
prevent, at finite temperature, the appearance of long-range order 
for infinite one-dimensional systems with short-range interactions \cite{mermi66_PRL}.

\end{appendix}

\bibliography{refs.bib}

\begin{thebibliography}{49}%
\makeatletter
\providecommand \@ifxundefined [1]{%
 \@ifx{#1\undefined}
}%
\providecommand \@ifnum [1]{%
 \ifnum #1\expandafter \@firstoftwo
 \else \expandafter \@secondoftwo
 \fi
}%
\providecommand \@ifx [1]{%
 \ifx #1\expandafter \@firstoftwo
 \else \expandafter \@secondoftwo
 \fi
}%
\providecommand \natexlab [1]{#1}%
\providecommand \enquote  [1]{``#1''}%
\providecommand \bibnamefont  [1]{#1}%
\providecommand \bibfnamefont [1]{#1}%
\providecommand \citenamefont [1]{#1}%
\providecommand \href@noop [0]{\@secondoftwo}%
\providecommand \href [0]{\begingroup \@sanitize@url \@href}%
\providecommand \@href[1]{\@@startlink{#1}\@@href}%
\providecommand \@@href[1]{\endgroup#1\@@endlink}%
\providecommand \@sanitize@url [0]{\catcode `\\12\catcode `\$12\catcode
  `\&12\catcode `\#12\catcode `\^12\catcode `\_12\catcode `\%12\relax}%
\providecommand \@@startlink[1]{}%
\providecommand \@@endlink[0]{}%
\providecommand \url  [0]{\begingroup\@sanitize@url \@url }%
\providecommand \@url [1]{\endgroup\@href {#1}{\urlprefix }}%
\providecommand \urlprefix  [0]{URL }%
\providecommand \Eprint [0]{\href }%
\providecommand \doibase [0]{http://dx.doi.org/}%
\providecommand \selectlanguage [0]{\@gobble}%
\providecommand \bibinfo  [0]{\@secondoftwo}%
\providecommand \bibfield  [0]{\@secondoftwo}%
\providecommand \translation [1]{[#1]}%
\providecommand \BibitemOpen [0]{}%
\providecommand \bibitemStop [0]{}%
\providecommand \bibitemNoStop [0]{.\EOS\space}%
\providecommand \EOS [0]{\spacefactor3000\relax}%
\providecommand \BibitemShut  [1]{\csname bibitem#1\endcsname}%
\let\auto@bib@innerbib\@empty
\bibitem [{\citenamefont {Suzuki}\ \emph {et~al.}(2012)\citenamefont {Suzuki},
  \citenamefont {Kawamura}, \citenamefont {Miyagawa}, \citenamefont
  {Garitaonandia}, \citenamefont {Yamamoto},\ and\ \citenamefont
  {Hori}}]{suzuk12_PRL}%
  \BibitemOpen
  \bibfield  {author} {\bibinfo {author} {\bibfnamefont {M.}~\bibnamefont
  {Suzuki}}, \bibinfo {author} {\bibfnamefont {N.}~\bibnamefont {Kawamura}},
  \bibinfo {author} {\bibfnamefont {H.}~\bibnamefont {Miyagawa}}, \bibinfo
  {author} {\bibfnamefont {J.~S.}\ \bibnamefont {Garitaonandia}}, \bibinfo
  {author} {\bibfnamefont {Y.}~\bibnamefont {Yamamoto}}, \ and\ \bibinfo
  {author} {\bibfnamefont {H.}~\bibnamefont {Hori}},\ }\bibfield  {title}
  {\enquote {\bibinfo {title} {Measurement of a {Pauli} and orbital
  paramagnetic state in bulk gold using {X}-ray magnetic circular dichroism
  spectroscopy},}\ }\href {\doibase 10.1103/PhysRevLett.108.047201} {\bibfield
  {journal} {\bibinfo  {journal} {Phys. Rev. Lett.}\ }\textbf {\bibinfo
  {volume} {108}},\ \bibinfo {pages} {047201} (\bibinfo {year}
  {2012})}\BibitemShut {NoStop}%
\bibitem [{\citenamefont {Hori}\ \emph {et~al.}(1999)\citenamefont {Hori},
  \citenamefont {Teranishi}, \citenamefont {Nakae}, \citenamefont {Seino},
  \citenamefont {Miyake},\ and\ \citenamefont {Yamada}}]{hori99_JPA}%
  \BibitemOpen
  \bibfield  {author} {\bibinfo {author} {\bibfnamefont {H.}~\bibnamefont
  {Hori}}, \bibinfo {author} {\bibfnamefont {T.}~\bibnamefont {Teranishi}},
  \bibinfo {author} {\bibfnamefont {Y.}~\bibnamefont {Nakae}}, \bibinfo
  {author} {\bibfnamefont {Y.}~\bibnamefont {Seino}}, \bibinfo {author}
  {\bibfnamefont {M.}~\bibnamefont {Miyake}}, \ and\ \bibinfo {author}
  {\bibfnamefont {S.}~\bibnamefont {Yamada}},\ }\bibfield  {title} {\enquote
  {\bibinfo {title} {{Anomalous magnetic polarization effect of Pd and Au
  nano-particles}},}\ }\href {\doibase 10.1016/S0375-9601(99)00742-2}
  {\bibfield  {journal} {\bibinfo  {journal} {Phys. Lett. A}\ }\textbf
  {\bibinfo {volume} {263}},\ \bibinfo {pages} {406} (\bibinfo {year}
  {1999})}\BibitemShut {NoStop}%
\bibitem [{\citenamefont {Nakae}\ \emph {et~al.}(2000)\citenamefont {Nakae},
  \citenamefont {Seino}, \citenamefont {Teranishi}, \citenamefont {Miyake},
  \citenamefont {Yamada},\ and\ \citenamefont {Hori}}]{nakae00_PhysicaB}%
  \BibitemOpen
  \bibfield  {author} {\bibinfo {author} {\bibfnamefont {Y.}~\bibnamefont
  {Nakae}}, \bibinfo {author} {\bibfnamefont {Y.}~\bibnamefont {Seino}},
  \bibinfo {author} {\bibfnamefont {T.}~\bibnamefont {Teranishi}}, \bibinfo
  {author} {\bibfnamefont {M.}~\bibnamefont {Miyake}}, \bibinfo {author}
  {\bibfnamefont {S.}~\bibnamefont {Yamada}}, \ and\ \bibinfo {author}
  {\bibfnamefont {H.}~\bibnamefont {Hori}},\ }\bibfield  {title} {\enquote
  {\bibinfo {title} {Anomalous spin polarization in {Pd and Au}
  nano-particles},}\ }\href {\doibase 10.1016/S0921-4526(99)02928-2} {\bibfield
   {journal} {\bibinfo  {journal} {Physica B}\ }\textbf {\bibinfo {volume}
  {284-288}},\ \bibinfo {pages} {1758} (\bibinfo {year} {2000})}\BibitemShut
  {NoStop}%
\bibitem [{\citenamefont {Hori}\ \emph {et~al.}(2004)\citenamefont {Hori},
  \citenamefont {Yamamoto}, \citenamefont {Iwamoto}, \citenamefont {Miura},
  \citenamefont {Teranishi},\ and\ \citenamefont {Miyake}}]{hori04_PRB}%
  \BibitemOpen
  \bibfield  {author} {\bibinfo {author} {\bibfnamefont {H.}~\bibnamefont
  {Hori}}, \bibinfo {author} {\bibfnamefont {Y.}~\bibnamefont {Yamamoto}},
  \bibinfo {author} {\bibfnamefont {T.}~\bibnamefont {Iwamoto}}, \bibinfo
  {author} {\bibfnamefont {T.}~\bibnamefont {Miura}}, \bibinfo {author}
  {\bibfnamefont {T.}~\bibnamefont {Teranishi}}, \ and\ \bibinfo {author}
  {\bibfnamefont {M.}~\bibnamefont {Miyake}},\ }\bibfield  {title} {\enquote
  {\bibinfo {title} {Diameter dependence of ferromagnetic spin moment in {Au}
  nanocrystals},}\ }\href {\doibase 10.1103/PhysRevB.69.174411} {\bibfield
  {journal} {\bibinfo  {journal} {Phys. Rev. B}\ }\textbf {\bibinfo {volume}
  {69}},\ \bibinfo {pages} {174411} (\bibinfo {year} {2004})}\BibitemShut
  {NoStop}%
\bibitem [{\citenamefont {Yamamoto}\ \emph {et~al.}(2004)\citenamefont
  {Yamamoto}, \citenamefont {Miura}, \citenamefont {Suzuki}, \citenamefont
  {Kawamura}, \citenamefont {Miyagawa}, \citenamefont {Nakamura}, \citenamefont
  {Kobayashi}, \citenamefont {Teranishi},\ and\ \citenamefont
  {Hori}}]{yamam04_PRL}%
  \BibitemOpen
  \bibfield  {author} {\bibinfo {author} {\bibfnamefont {Y.}~\bibnamefont
  {Yamamoto}}, \bibinfo {author} {\bibfnamefont {T.}~\bibnamefont {Miura}},
  \bibinfo {author} {\bibfnamefont {M.}~\bibnamefont {Suzuki}}, \bibinfo
  {author} {\bibfnamefont {N.}~\bibnamefont {Kawamura}}, \bibinfo {author}
  {\bibfnamefont {H.}~\bibnamefont {Miyagawa}}, \bibinfo {author}
  {\bibfnamefont {T.}~\bibnamefont {Nakamura}}, \bibinfo {author}
  {\bibfnamefont {K.}~\bibnamefont {Kobayashi}}, \bibinfo {author}
  {\bibfnamefont {T.}~\bibnamefont {Teranishi}}, \ and\ \bibinfo {author}
  {\bibfnamefont {H.}~\bibnamefont {Hori}},\ }\bibfield  {title} {\enquote
  {\bibinfo {title} {Direct observation of ferromagnetic spin polarization in
  gold nanoparticles},}\ }\href {\doibase 10.1103/PhysRevLett.93.116801}
  {\bibfield  {journal} {\bibinfo  {journal} {Phys. Rev. Lett.}\ }\textbf
  {\bibinfo {volume} {93}},\ \bibinfo {pages} {116801} (\bibinfo {year}
  {2004})}\BibitemShut {NoStop}%
\bibitem [{\citenamefont {Guerrero}\ \emph
  {et~al.}(2008{\natexlab{a}})\citenamefont {Guerrero}, \citenamefont
  {Mu{\~n}oz-M{\'a}rquez}, \citenamefont {Fern{\'a}ndez-Pinel}, \citenamefont
  {Crespo}, \citenamefont {Hernando},\ and\ \citenamefont
  {Fern{\'a}ndez}}]{guerr08}%
  \BibitemOpen
  \bibfield  {author} {\bibinfo {author} {\bibfnamefont {E.}~\bibnamefont
  {Guerrero}}, \bibinfo {author} {\bibfnamefont {M.~A.}\ \bibnamefont
  {Mu{\~n}oz-M{\'a}rquez}}, \bibinfo {author} {\bibfnamefont {E.}~\bibnamefont
  {Fern{\'a}ndez-Pinel}}, \bibinfo {author} {\bibfnamefont {P.}~\bibnamefont
  {Crespo}}, \bibinfo {author} {\bibfnamefont {A.}~\bibnamefont {Hernando}}, \
  and\ \bibinfo {author} {\bibfnamefont {A.}~\bibnamefont {Fern{\'a}ndez}},\
  }\bibfield  {title} {\enquote {\bibinfo {title} {Electronic structure,
  magnetic properties, and microstructural analysis of thiol-functionalized
  {Au} nanoparticles: role of chemical and structural parameters in the
  ferromagnetic behaviour},}\ }\href {\doibase 10.1007/s11051-008-9445-5}
  {\bibfield  {journal} {\bibinfo  {journal} {J. Nanopart. Res.}\ }\textbf
  {\bibinfo {volume} {10}},\ \bibinfo {pages} {179} (\bibinfo {year}
  {2008}{\natexlab{a}})}\BibitemShut {NoStop}%
\bibitem [{\citenamefont {Guerrero}\ \emph
  {et~al.}(2008{\natexlab{b}})\citenamefont {Guerrero}, \citenamefont
  {Mu{\~{n}}oz-M{\'{a}}rquez}, \citenamefont {Garc{\'{\i}}a}, \citenamefont
  {Crespo}, \citenamefont {Fern{\'{a}}ndez-Pinel}, \citenamefont {Hernando},\
  and\ \citenamefont {Fern{\'{a}}ndez}}]{guerr08_Nanotechnology}%
  \BibitemOpen
  \bibfield  {author} {\bibinfo {author} {\bibfnamefont {E.}~\bibnamefont
  {Guerrero}}, \bibinfo {author} {\bibfnamefont {M.~A.}\ \bibnamefont
  {Mu{\~{n}}oz-M{\'{a}}rquez}}, \bibinfo {author} {\bibfnamefont {M.~A.}\
  \bibnamefont {Garc{\'{\i}}a}}, \bibinfo {author} {\bibfnamefont
  {P.}~\bibnamefont {Crespo}}, \bibinfo {author} {\bibfnamefont
  {E.}~\bibnamefont {Fern{\'{a}}ndez-Pinel}}, \bibinfo {author} {\bibfnamefont
  {A.}~\bibnamefont {Hernando}}, \ and\ \bibinfo {author} {\bibfnamefont
  {A.}~\bibnamefont {Fern{\'{a}}ndez}},\ }\bibfield  {title} {\enquote
  {\bibinfo {title} {Surface plasmon resonance and magnetism of thiol-capped
  gold nanoparticles},}\ }\href {\doibase 10.1088/0957-4484/19/17/175701}
  {\bibfield  {journal} {\bibinfo  {journal} {Nanotechnology}\ }\textbf
  {\bibinfo {volume} {19}},\ \bibinfo {pages} {175701} (\bibinfo {year}
  {2008}{\natexlab{b}})}\BibitemShut {NoStop}%
\bibitem [{\citenamefont {Bartolom\'e}\ \emph {et~al.}(2012)\citenamefont
  {Bartolom\'e}, \citenamefont {Bartolom\'e}, \citenamefont {Garc\'{\i}a},
  \citenamefont {Figueroa}, \citenamefont {Repoll\'es}, \citenamefont
  {Mart\'{\i}nez-P\'erez}, \citenamefont {Luis}, \citenamefont {Mag\'en},
  \citenamefont {Selenska-Pobell}, \citenamefont {Pobell}, \citenamefont
  {Reitz}, \citenamefont {Sch\"onemann}, \citenamefont {Herrmannsd\"orfer},
  \citenamefont {Merroun}, \citenamefont {Geissler}, \citenamefont {Wilhelm},\
  and\ \citenamefont {Rogalev}}]{barto12_PRL}%
  \BibitemOpen
  \bibfield  {author} {\bibinfo {author} {\bibfnamefont {J.}~\bibnamefont
  {Bartolom\'e}}, \bibinfo {author} {\bibfnamefont {F.}~\bibnamefont
  {Bartolom\'e}}, \bibinfo {author} {\bibfnamefont {L.~M.}\ \bibnamefont
  {Garc\'{\i}a}}, \bibinfo {author} {\bibfnamefont {A.~I.}\ \bibnamefont
  {Figueroa}}, \bibinfo {author} {\bibfnamefont {A.}~\bibnamefont
  {Repoll\'es}}, \bibinfo {author} {\bibfnamefont {M.~J.}\ \bibnamefont
  {Mart\'{\i}nez-P\'erez}}, \bibinfo {author} {\bibfnamefont {F.}~\bibnamefont
  {Luis}}, \bibinfo {author} {\bibfnamefont {C.}~\bibnamefont {Mag\'en}},
  \bibinfo {author} {\bibfnamefont {S.}~\bibnamefont {Selenska-Pobell}},
  \bibinfo {author} {\bibfnamefont {F.}~\bibnamefont {Pobell}}, \bibinfo
  {author} {\bibfnamefont {T.}~\bibnamefont {Reitz}}, \bibinfo {author}
  {\bibfnamefont {R.}~\bibnamefont {Sch\"onemann}}, \bibinfo {author}
  {\bibfnamefont {T.}~\bibnamefont {Herrmannsd\"orfer}}, \bibinfo {author}
  {\bibfnamefont {M.}~\bibnamefont {Merroun}}, \bibinfo {author} {\bibfnamefont
  {A.}~\bibnamefont {Geissler}}, \bibinfo {author} {\bibfnamefont
  {F.}~\bibnamefont {Wilhelm}}, \ and\ \bibinfo {author} {\bibfnamefont
  {A.}~\bibnamefont {Rogalev}},\ }\bibfield  {title} {\enquote {\bibinfo
  {title} {Strong paramagnetism of gold nanoparticles deposited on a sulfolobus
  acidocaldarius {$S$} layer},}\ }\href {\doibase
  10.1103/PhysRevLett.109.247203} {\bibfield  {journal} {\bibinfo  {journal}
  {Phys. Rev. Lett.}\ }\textbf {\bibinfo {volume} {109}},\ \bibinfo {pages}
  {247203} (\bibinfo {year} {2012})}\BibitemShut {NoStop}%
\bibitem [{\citenamefont {Agrachev}\ \emph {et~al.}(2017)\citenamefont
  {Agrachev}, \citenamefont {Antonello}, \citenamefont {Dainese}, \citenamefont
  {Ruzzi}, \citenamefont {Zoleo}, \citenamefont {Apr{\`a}}, \citenamefont
  {Govind}, \citenamefont {Fortunelli}, \citenamefont {Sementa},\ and\
  \citenamefont {Maran}}]{agrac17_ACSOmega}%
  \BibitemOpen
  \bibfield  {author} {\bibinfo {author} {\bibfnamefont {M.}~\bibnamefont
  {Agrachev}}, \bibinfo {author} {\bibfnamefont {S.}~\bibnamefont {Antonello}},
  \bibinfo {author} {\bibfnamefont {T.}~\bibnamefont {Dainese}}, \bibinfo
  {author} {\bibfnamefont {M.}~\bibnamefont {Ruzzi}}, \bibinfo {author}
  {\bibfnamefont {A.}~\bibnamefont {Zoleo}}, \bibinfo {author} {\bibfnamefont
  {E.}~\bibnamefont {Apr{\`a}}}, \bibinfo {author} {\bibfnamefont
  {N.}~\bibnamefont {Govind}}, \bibinfo {author} {\bibfnamefont
  {A.}~\bibnamefont {Fortunelli}}, \bibinfo {author} {\bibfnamefont
  {L.}~\bibnamefont {Sementa}}, \ and\ \bibinfo {author} {\bibfnamefont
  {F.}~\bibnamefont {Maran}},\ }\bibfield  {title} {\enquote {\bibinfo {title}
  {Magnetic ordering in gold nanoclusters},}\ }\href {\doibase
  10.1021/acsomega.7b00472} {\bibfield  {journal} {\bibinfo  {journal} {ACS
  Omega}\ }\textbf {\bibinfo {volume} {2}},\ \bibinfo {pages} {2607} (\bibinfo
  {year} {2017})}\BibitemShut {NoStop}%
\bibitem [{\citenamefont {Crespo}\ \emph {et~al.}(2004)\citenamefont {Crespo},
  \citenamefont {Litr\'an}, \citenamefont {Rojas}, \citenamefont {Multigner},
  \citenamefont {de~la Fuente}, \citenamefont {S\'anchez-L\'opez},
  \citenamefont {Garc\'{\i}a}, \citenamefont {Hernando}, \citenamefont
  {Penad\'es},\ and\ \citenamefont {Fern\'andez}}]{cresp04_PRL}%
  \BibitemOpen
  \bibfield  {author} {\bibinfo {author} {\bibfnamefont {P.}~\bibnamefont
  {Crespo}}, \bibinfo {author} {\bibfnamefont {R.}~\bibnamefont {Litr\'an}},
  \bibinfo {author} {\bibfnamefont {T.~C.}\ \bibnamefont {Rojas}}, \bibinfo
  {author} {\bibfnamefont {M.}~\bibnamefont {Multigner}}, \bibinfo {author}
  {\bibfnamefont {J.~M.}\ \bibnamefont {de~la Fuente}}, \bibinfo {author}
  {\bibfnamefont {J.~C.}\ \bibnamefont {S\'anchez-L\'opez}}, \bibinfo {author}
  {\bibfnamefont {M.~A.}\ \bibnamefont {Garc\'{\i}a}}, \bibinfo {author}
  {\bibfnamefont {A.}~\bibnamefont {Hernando}}, \bibinfo {author}
  {\bibfnamefont {S.}~\bibnamefont {Penad\'es}}, \ and\ \bibinfo {author}
  {\bibfnamefont {A.}~\bibnamefont {Fern\'andez}},\ }\bibfield  {title}
  {\enquote {\bibinfo {title} {Permanent magnetism, magnetic anisotropy, and
  hysteresis of thiol-capped gold nanoparticles},}\ }\href {\doibase
  10.1103/PhysRevLett.93.087204} {\bibfield  {journal} {\bibinfo  {journal}
  {Phys. Rev. Lett.}\ }\textbf {\bibinfo {volume} {93}},\ \bibinfo {pages}
  {087204} (\bibinfo {year} {2004})}\BibitemShut {NoStop}%
\bibitem [{\citenamefont {Dutta}\ \emph {et~al.}(2007)\citenamefont {Dutta},
  \citenamefont {Pal}, \citenamefont {Seehra}, \citenamefont {Anand},\ and\
  \citenamefont {Roberts}}]{dutta07_APL}%
  \BibitemOpen
  \bibfield  {author} {\bibinfo {author} {\bibfnamefont {P.}~\bibnamefont
  {Dutta}}, \bibinfo {author} {\bibfnamefont {S.}~\bibnamefont {Pal}}, \bibinfo
  {author} {\bibfnamefont {M.~S.}\ \bibnamefont {Seehra}}, \bibinfo {author}
  {\bibfnamefont {M.}~\bibnamefont {Anand}}, \ and\ \bibinfo {author}
  {\bibfnamefont {C.~B.}\ \bibnamefont {Roberts}},\ }\bibfield  {title}
  {\enquote {\bibinfo {title} {Magnetism in dodecanethiol-capped gold
  nanoparticles: Role of size and capping agent},}\ }\href {\doibase
  10.1063/1.2740577} {\bibfield  {journal} {\bibinfo  {journal} {Appl. Phys.
  Lett.}\ }\textbf {\bibinfo {volume} {90}},\ \bibinfo {pages} {213102}
  (\bibinfo {year} {2007})}\BibitemShut {NoStop}%
\bibitem [{\citenamefont {van Rhee}\ \emph {et~al.}(2013)\citenamefont {van
  Rhee}, \citenamefont {Zijlstra}, \citenamefont {Verhagen}, \citenamefont
  {Aarts}, \citenamefont {Katsnelson}, \citenamefont {Maan}, \citenamefont
  {Orrit},\ and\ \citenamefont {Christianen}}]{rhee13_PRL}%
  \BibitemOpen
  \bibfield  {author} {\bibinfo {author} {\bibfnamefont {P.~G.}\ \bibnamefont
  {van Rhee}}, \bibinfo {author} {\bibfnamefont {P.}~\bibnamefont {Zijlstra}},
  \bibinfo {author} {\bibfnamefont {T.~G.~A.}\ \bibnamefont {Verhagen}},
  \bibinfo {author} {\bibfnamefont {J.}~\bibnamefont {Aarts}}, \bibinfo
  {author} {\bibfnamefont {M.~I.}\ \bibnamefont {Katsnelson}}, \bibinfo
  {author} {\bibfnamefont {J.~C.}\ \bibnamefont {Maan}}, \bibinfo {author}
  {\bibfnamefont {M.}~\bibnamefont {Orrit}}, \ and\ \bibinfo {author}
  {\bibfnamefont {P.~C.~M.}\ \bibnamefont {Christianen}},\ }\bibfield  {title}
  {\enquote {\bibinfo {title} {Giant magnetic susceptibility of gold nanorods
  detected by magnetic alignment},}\ }\href {\doibase
  10.1103/PhysRevLett.111.127202} {\bibfield  {journal} {\bibinfo  {journal}
  {Phys. Rev. Lett.}\ }\textbf {\bibinfo {volume} {111}},\ \bibinfo {pages}
  {127202} (\bibinfo {year} {2013})}\BibitemShut {NoStop}%
\bibitem [{\citenamefont {Crespo}\ \emph {et~al.}(2006)\citenamefont {Crespo},
  \citenamefont {Garc\'{\i}a}, \citenamefont {Fern\'andez~Pinel}, \citenamefont
  {Multigner}, \citenamefont {Alc\'antara}, \citenamefont {de~la Fuente},
  \citenamefont {Penad\'es},\ and\ \citenamefont {Hernando}}]{cresp06_PRL}%
  \BibitemOpen
  \bibfield  {author} {\bibinfo {author} {\bibfnamefont {P.}~\bibnamefont
  {Crespo}}, \bibinfo {author} {\bibfnamefont {M.~A.}\ \bibnamefont
  {Garc\'{\i}a}}, \bibinfo {author} {\bibfnamefont {E.}~\bibnamefont
  {Fern\'andez~Pinel}}, \bibinfo {author} {\bibfnamefont {M.}~\bibnamefont
  {Multigner}}, \bibinfo {author} {\bibfnamefont {D.}~\bibnamefont
  {Alc\'antara}}, \bibinfo {author} {\bibfnamefont {J.~M.}\ \bibnamefont {de~la
  Fuente}}, \bibinfo {author} {\bibfnamefont {S.}~\bibnamefont {Penad\'es}}, \
  and\ \bibinfo {author} {\bibfnamefont {A.}~\bibnamefont {Hernando}},\
  }\bibfield  {title} {\enquote {\bibinfo {title} {Fe impurities weaken the
  ferromagnetic behavior in {Au} nanoparticles},}\ }\href {\doibase
  10.1103/PhysRevLett.97.177203} {\bibfield  {journal} {\bibinfo  {journal}
  {Phys. Rev. Lett.}\ }\textbf {\bibinfo {volume} {97}},\ \bibinfo {pages}
  {177203} (\bibinfo {year} {2006})}\BibitemShut {NoStop}%
\bibitem [{\citenamefont {Donnio}\ \emph {et~al.}(2007)\citenamefont {Donnio},
  \citenamefont {García-Vázquez}, \citenamefont {Gallani}, \citenamefont
  {Guillon},\ and\ \citenamefont {Terazzi}}]{donni07_AdvMater}%
  \BibitemOpen
  \bibfield  {author} {\bibinfo {author} {\bibfnamefont {B.}~\bibnamefont
  {Donnio}}, \bibinfo {author} {\bibfnamefont {P.}~\bibnamefont
  {García-Vázquez}}, \bibinfo {author} {\bibfnamefont {J.-L.}\ \bibnamefont
  {Gallani}}, \bibinfo {author} {\bibfnamefont {D.}~\bibnamefont {Guillon}}, \
  and\ \bibinfo {author} {\bibfnamefont {E.}~\bibnamefont {Terazzi}},\
  }\bibfield  {title} {\enquote {\bibinfo {title} {Dendronized ferromagnetic
  gold nanoparticles self-organized in a thermotropic cubic phase},}\ }\href
  {\doibase 10.1002/adma.200701252} {\bibfield  {journal} {\bibinfo  {journal}
  {Adv. Mater.}\ }\textbf {\bibinfo {volume} {19}},\ \bibinfo {pages} {3534}
  (\bibinfo {year} {2007})}\BibitemShut {NoStop}%
\bibitem [{\citenamefont {Garitaonandia}\ \emph {et~al.}(2008)\citenamefont
  {Garitaonandia}, \citenamefont {Insausti}, \citenamefont {Goikolea},
  \citenamefont {Suzuki}, \citenamefont {Cashion}, \citenamefont {Kawamura},
  \citenamefont {Ohsawa}, \citenamefont {Gil~de Muro}, \citenamefont {Suzuki},
  \citenamefont {Plazaola},\ and\ \citenamefont {Rojo}}]{garit08_NL}%
  \BibitemOpen
  \bibfield  {author} {\bibinfo {author} {\bibfnamefont {J.~S.}\ \bibnamefont
  {Garitaonandia}}, \bibinfo {author} {\bibfnamefont {M.}~\bibnamefont
  {Insausti}}, \bibinfo {author} {\bibfnamefont {E.}~\bibnamefont {Goikolea}},
  \bibinfo {author} {\bibfnamefont {M.}~\bibnamefont {Suzuki}}, \bibinfo
  {author} {\bibfnamefont {J.~D.}\ \bibnamefont {Cashion}}, \bibinfo {author}
  {\bibfnamefont {N.}~\bibnamefont {Kawamura}}, \bibinfo {author}
  {\bibfnamefont {H.}~\bibnamefont {Ohsawa}}, \bibinfo {author} {\bibfnamefont
  {I.}~\bibnamefont {Gil~de Muro}}, \bibinfo {author} {\bibfnamefont
  {K.}~\bibnamefont {Suzuki}}, \bibinfo {author} {\bibfnamefont
  {F.}~\bibnamefont {Plazaola}}, \ and\ \bibinfo {author} {\bibfnamefont
  {T.}~\bibnamefont {Rojo}},\ }\bibfield  {title} {\enquote {\bibinfo {title}
  {Chemically induced permanent magnetism in {Au, Ag, and Cu} nanoparticles:
  localization of the magnetism by element selective techniques},}\ }\href
  {\doibase 10.1021/nl073129g} {\bibfield  {journal} {\bibinfo  {journal} {Nano
  Lett.}\ }\textbf {\bibinfo {volume} {8}},\ \bibinfo {pages} {661} (\bibinfo
  {year} {2008})}\BibitemShut {NoStop}%
\bibitem [{\citenamefont {de~la Venta}\ \emph {et~al.}(2009)\citenamefont
  {de~la Venta}, \citenamefont {Bouzas}, \citenamefont {Pucci}, \citenamefont
  {Laguna-Marco}, \citenamefont {Haskel}, \citenamefont {te~Velthuis},
  \citenamefont {Hoffmann}, \citenamefont {Lal}, \citenamefont {Bleuel},
  \citenamefont {Ruggeri}, \citenamefont {de~Juli{\'a}n~Fern{\'a}ndez},\ and\
  \citenamefont {Garc{\'i}a}}]{venta09}%
  \BibitemOpen
  \bibfield  {author} {\bibinfo {author} {\bibfnamefont {J.}~\bibnamefont
  {de~la Venta}}, \bibinfo {author} {\bibfnamefont {V.}~\bibnamefont {Bouzas}},
  \bibinfo {author} {\bibfnamefont {A.}~\bibnamefont {Pucci}}, \bibinfo
  {author} {\bibfnamefont {M.~A.}\ \bibnamefont {Laguna-Marco}}, \bibinfo
  {author} {\bibfnamefont {D.}~\bibnamefont {Haskel}}, \bibinfo {author}
  {\bibfnamefont {S.~G.~E.}\ \bibnamefont {te~Velthuis}}, \bibinfo {author}
  {\bibfnamefont {A.}~\bibnamefont {Hoffmann}}, \bibinfo {author}
  {\bibfnamefont {J.}~\bibnamefont {Lal}}, \bibinfo {author} {\bibfnamefont
  {M.}~\bibnamefont {Bleuel}}, \bibinfo {author} {\bibfnamefont
  {G.}~\bibnamefont {Ruggeri}}, \bibinfo {author} {\bibfnamefont
  {C.}~\bibnamefont {de~Juli{\'a}n~Fern{\'a}ndez}}, \ and\ \bibinfo {author}
  {\bibfnamefont {M.~A.}\ \bibnamefont {Garc{\'i}a}},\ }\bibfield  {title}
  {\enquote {\bibinfo {title} {X-ray magnetic circular dichroism and small
  angle neutron scattering studies of thiol capped gold nanoparticles},}\
  }\href {\doibase doi:10.1166/jnn.2009.1877} {\bibfield  {journal} {\bibinfo
  {journal} {J. Nanosci. Nanotechnol.}\ }\textbf {\bibinfo {volume} {9}},\
  \bibinfo {pages} {6434} (\bibinfo {year} {2009})}\BibitemShut {NoStop}%
\bibitem [{\citenamefont {Donnio}\ \emph {et~al.}(2010)\citenamefont {Donnio},
  \citenamefont {Derory}, \citenamefont {Terazzi}, \citenamefont {Drillon},
  \citenamefont {Guillon},\ and\ \citenamefont {Gallani}}]{donni10_SM}%
  \BibitemOpen
  \bibfield  {author} {\bibinfo {author} {\bibfnamefont {B.}~\bibnamefont
  {Donnio}}, \bibinfo {author} {\bibfnamefont {A.}~\bibnamefont {Derory}},
  \bibinfo {author} {\bibfnamefont {E.}~\bibnamefont {Terazzi}}, \bibinfo
  {author} {\bibfnamefont {M.}~\bibnamefont {Drillon}}, \bibinfo {author}
  {\bibfnamefont {D.}~\bibnamefont {Guillon}}, \ and\ \bibinfo {author}
  {\bibfnamefont {J.-L.}\ \bibnamefont {Gallani}},\ }\bibfield  {title}
  {\enquote {\bibinfo {title} {Very slow high-temperature relaxation of the
  remnant magnetic moment in 2 nm mesomorphic gold nanoparticles},}\ }\href
  {\doibase 10.1039/B918602F} {\bibfield  {journal} {\bibinfo  {journal} {Soft
  Matter}\ }\textbf {\bibinfo {volume} {6}},\ \bibinfo {pages} {965} (\bibinfo
  {year} {2010})}\BibitemShut {NoStop}%
\bibitem [{\citenamefont {Maitra}\ \emph {et~al.}(2011)\citenamefont {Maitra},
  \citenamefont {Das}, \citenamefont {Kumar}, \citenamefont {Sundaresan},\ and\
  \citenamefont {Rao}}]{maitr11_CPC}%
  \BibitemOpen
  \bibfield  {author} {\bibinfo {author} {\bibfnamefont {U.}~\bibnamefont
  {Maitra}}, \bibinfo {author} {\bibfnamefont {B.}~\bibnamefont {Das}},
  \bibinfo {author} {\bibfnamefont {N.}~\bibnamefont {Kumar}}, \bibinfo
  {author} {\bibfnamefont {A.}~\bibnamefont {Sundaresan}}, \ and\ \bibinfo
  {author} {\bibfnamefont {C.~N.~R.}\ \bibnamefont {Rao}},\ }\bibfield  {title}
  {\enquote {\bibinfo {title} {Ferromagnetism exhibited by nanoparticles of
  noble metals},}\ }\href {\doibase 10.1002/cphc.201100121} {\bibfield
  {journal} {\bibinfo  {journal} {ChemPhysChem}\ }\textbf {\bibinfo {volume}
  {12}},\ \bibinfo {pages} {2322} (\bibinfo {year} {2011})}\BibitemShut
  {NoStop}%
\bibitem [{\citenamefont {Gréget}\ \emph {et~al.}(2012)\citenamefont
  {Gréget}, \citenamefont {Nealon}, \citenamefont {Vileno}, \citenamefont
  {Turek}, \citenamefont {Mény}, \citenamefont {Ott}, \citenamefont {Derory},
  \citenamefont {Voirin}, \citenamefont {Rivière}, \citenamefont {Rogalev},
  \citenamefont {Wilhelm}, \citenamefont {Joly}, \citenamefont {Knafo},
  \citenamefont {Ballon}, \citenamefont {Terazzi}, \citenamefont {Kappler},
  \citenamefont {Donnio},\ and\ \citenamefont {Gallani}}]{grege12_CPC}%
  \BibitemOpen
  \bibfield  {author} {\bibinfo {author} {\bibfnamefont {R.}~\bibnamefont
  {Gréget}}, \bibinfo {author} {\bibfnamefont {G.~L.}\ \bibnamefont {Nealon}},
  \bibinfo {author} {\bibfnamefont {B.}~\bibnamefont {Vileno}}, \bibinfo
  {author} {\bibfnamefont {P.}~\bibnamefont {Turek}}, \bibinfo {author}
  {\bibfnamefont {C.}~\bibnamefont {Mény}}, \bibinfo {author} {\bibfnamefont
  {F.}~\bibnamefont {Ott}}, \bibinfo {author} {\bibfnamefont {A.}~\bibnamefont
  {Derory}}, \bibinfo {author} {\bibfnamefont {E.}~\bibnamefont {Voirin}},
  \bibinfo {author} {\bibfnamefont {E.}~\bibnamefont {Rivière}}, \bibinfo
  {author} {\bibfnamefont {A.}~\bibnamefont {Rogalev}}, \bibinfo {author}
  {\bibfnamefont {F.}~\bibnamefont {Wilhelm}}, \bibinfo {author} {\bibfnamefont
  {L.}~\bibnamefont {Joly}}, \bibinfo {author} {\bibfnamefont {W.}~\bibnamefont
  {Knafo}}, \bibinfo {author} {\bibfnamefont {G.}~\bibnamefont {Ballon}},
  \bibinfo {author} {\bibfnamefont {E.}~\bibnamefont {Terazzi}}, \bibinfo
  {author} {\bibfnamefont {J.-P.}\ \bibnamefont {Kappler}}, \bibinfo {author}
  {\bibfnamefont {B.}~\bibnamefont {Donnio}}, \ and\ \bibinfo {author}
  {\bibfnamefont {J.-L.}\ \bibnamefont {Gallani}},\ }\bibfield  {title}
  {\enquote {\bibinfo {title} {Magnetic properties of gold nanoparticles: A
  room-temperature quantum effect},}\ }\href {\doibase 10.1002/cphc.201200394}
  {\bibfield  {journal} {\bibinfo  {journal} {ChemPhysChem}\ }\textbf {\bibinfo
  {volume} {13}},\ \bibinfo {pages} {3092} (\bibinfo {year}
  {2012})}\BibitemShut {NoStop}%
\bibitem [{\citenamefont {Nealon}\ \emph {et~al.}(2012)\citenamefont {Nealon},
  \citenamefont {Donnio}, \citenamefont {Greget}, \citenamefont {Kappler},
  \citenamefont {Terazzi},\ and\ \citenamefont {Gallani}}]{nealo12_Nanoscale}%
  \BibitemOpen
  \bibfield  {author} {\bibinfo {author} {\bibfnamefont {G.~L.}\ \bibnamefont
  {Nealon}}, \bibinfo {author} {\bibfnamefont {B.}~\bibnamefont {Donnio}},
  \bibinfo {author} {\bibfnamefont {R.}~\bibnamefont {Greget}}, \bibinfo
  {author} {\bibfnamefont {J.-P.}\ \bibnamefont {Kappler}}, \bibinfo {author}
  {\bibfnamefont {E.}~\bibnamefont {Terazzi}}, \ and\ \bibinfo {author}
  {\bibfnamefont {J.-L.}\ \bibnamefont {Gallani}},\ }\bibfield  {title}
  {\enquote {\bibinfo {title} {Magnetism in gold nanoparticles},}\ }\href
  {\doibase 10.1039/C2NR30640A} {\bibfield  {journal} {\bibinfo  {journal}
  {Nanoscale}\ }\textbf {\bibinfo {volume} {4}},\ \bibinfo {pages} {5244}
  (\bibinfo {year} {2012})}\BibitemShut {NoStop}%
\bibitem [{\citenamefont {G\'omez~Viloria}\ \emph {et~al.}(2018)\citenamefont
  {G\'omez~Viloria}, \citenamefont {Weick}, \citenamefont {Weinmann},\ and\
  \citenamefont {Jalabert}}]{gomez18_PRB}%
  \BibitemOpen
  \bibfield  {author} {\bibinfo {author} {\bibfnamefont {M.}~\bibnamefont
  {G\'omez~Viloria}}, \bibinfo {author} {\bibfnamefont {G.}~\bibnamefont
  {Weick}}, \bibinfo {author} {\bibfnamefont {D.}~\bibnamefont {Weinmann}}, \
  and\ \bibinfo {author} {\bibfnamefont {R.~A.}\ \bibnamefont {Jalabert}},\
  }\bibfield  {title} {\enquote {\bibinfo {title} {Orbital magnetism in
  ensembles of gold nanoparticles},}\ }\href {\doibase
  10.1103/PhysRevB.98.195417} {\bibfield  {journal} {\bibinfo  {journal} {Phys.
  Rev. B}\ }\textbf {\bibinfo {volume} {98}},\ \bibinfo {pages} {195417}
  (\bibinfo {year} {2018})}\BibitemShut {NoStop}%
\bibitem [{\citenamefont {Landau}\ and\ \citenamefont
  {Lifshitz}(1985)}]{landau_statphys}%
  \BibitemOpen
  \bibfield  {author} {\bibinfo {author} {\bibfnamefont {L.~D.}\ \bibnamefont
  {Landau}}\ and\ \bibinfo {author} {\bibfnamefont {E.~M.}\ \bibnamefont
  {Lifshitz}},\ }\href@noop {} {\emph {\bibinfo {title} {Statistical
  Physics}}}\ (\bibinfo  {publisher} {Pergamon},\ \bibinfo {address} {Oxford},\
  \bibinfo {year} {1985})\BibitemShut {NoStop}%
\bibitem [{\citenamefont {Richter}\ \emph {et~al.}(1996)\citenamefont
  {Richter}, \citenamefont {Ullmo},\ and\ \citenamefont
  {Jalabert}}]{richt96_PhysRep}%
  \BibitemOpen
  \bibfield  {author} {\bibinfo {author} {\bibfnamefont {K.}~\bibnamefont
  {Richter}}, \bibinfo {author} {\bibfnamefont {D.}~\bibnamefont {Ullmo}}, \
  and\ \bibinfo {author} {\bibfnamefont {R.~A.}\ \bibnamefont {Jalabert}},\
  }\bibfield  {title} {\enquote {\bibinfo {title} {Orbital magnetism in the
  ballistic regime: geometrical effects},}\ }\href {\doibase
  10.1016/0370-1573(96)00010-5} {\bibfield  {journal} {\bibinfo  {journal}
  {Phys. Rep.}\ }\textbf {\bibinfo {volume} {276}},\ \bibinfo {pages} {1}
  (\bibinfo {year} {1996})}\BibitemShut {NoStop}%
\bibitem [{\citenamefont {van Ruitenbeek}\ and\ \citenamefont {van
  Leeuwen}(1991)}]{ruite91_PRL}%
  \BibitemOpen
  \bibfield  {author} {\bibinfo {author} {\bibfnamefont {J.~M.}\ \bibnamefont
  {van Ruitenbeek}}\ and\ \bibinfo {author} {\bibfnamefont {D.~A.}\
  \bibnamefont {van Leeuwen}},\ }\bibfield  {title} {\enquote {\bibinfo {title}
  {Model calculation of size effects in orbital magnetism},}\ }\href {\doibase
  10.1103/PhysRevLett.67.640} {\bibfield  {journal} {\bibinfo  {journal} {Phys.
  Rev. Lett.}\ }\textbf {\bibinfo {volume} {67}},\ \bibinfo {pages} {640}
  (\bibinfo {year} {1991})}\BibitemShut {NoStop}%
\bibitem [{\citenamefont {van Ruitenbeek}\ and\ \citenamefont {van
  Leeuwen}(1993)}]{ruite93_MPLB}%
  \BibitemOpen
  \bibfield  {author} {\bibinfo {author} {\bibfnamefont {J.~M.}\ \bibnamefont
  {van Ruitenbeek}}\ and\ \bibinfo {author} {\bibfnamefont {D.~A.}\
  \bibnamefont {van Leeuwen}},\ }\bibfield  {title} {\enquote {\bibinfo {title}
  {Size effects in orbital magnetism},}\ }\href {\doibase
  10.1142/S0217984993001053} {\bibfield  {journal} {\bibinfo  {journal} {Mod.
  Phys. Lett. B}\ }\textbf {\bibinfo {volume} {07}},\ \bibinfo {pages} {1053}
  (\bibinfo {year} {1993})}\BibitemShut {NoStop}%
\bibitem [{\citenamefont {van Leeuwen}(1993)}]{leuwe93_PhD}%
  \BibitemOpen
  \bibfield  {author} {\bibinfo {author} {\bibfnamefont {D.~A.}\ \bibnamefont
  {van Leeuwen}},\ }\href@noop {} {\emph {\bibinfo {title} {Magnetic Moments in
  Metacluster Molecules}}}\ (\bibinfo  {publisher} {PhD thesis, University of
  Leiden, unpublished},\ \bibinfo {year} {1993})\BibitemShut {NoStop}%
\bibitem [{\citenamefont {Imry}(2015)}]{imry15_PRB}%
  \BibitemOpen
  \bibfield  {author} {\bibinfo {author} {\bibfnamefont {Y.}~\bibnamefont
  {Imry}},\ }\bibfield  {title} {\enquote {\bibinfo {title} {Superconducting
  fluctuations and large diamagnetism of low-${T}_\mathrm{c}$ nanoparticles},}\
  }\href {\doibase 10.1103/PhysRevB.91.104503} {\bibfield  {journal} {\bibinfo
  {journal} {Phys. Rev. B}\ }\textbf {\bibinfo {volume} {91}},\ \bibinfo
  {pages} {104503} (\bibinfo {year} {2015})}\BibitemShut {NoStop}%
\bibitem [{\citenamefont {Aslamazov}\ and\ \citenamefont
  {Larkin}(1975)}]{aslam75_JETP}%
  \BibitemOpen
  \bibfield  {author} {\bibinfo {author} {\bibfnamefont {L.~G.}\ \bibnamefont
  {Aslamazov}}\ and\ \bibinfo {author} {\bibfnamefont {A.~I.}\ \bibnamefont
  {Larkin}},\ }\bibfield  {title} {\enquote {\bibinfo {title}
  {Fluctuation-induced magnetic susceptibility of superconductors and normal
  metals},}\ }\href@noop {} {\bibfield  {journal} {\bibinfo  {journal} {Sov.
  Phys. JETP}\ }\textbf {\bibinfo {volume} {40}},\ \bibinfo {pages} {321}
  (\bibinfo {year} {1975})}\BibitemShut {NoStop}%
\bibitem [{\citenamefont {Murzaliev}\ \emph {et~al.}(2019)\citenamefont
  {Murzaliev}, \citenamefont {Titov},\ and\ \citenamefont
  {Katsnelson}}]{murza19_PRB}%
  \BibitemOpen
  \bibfield  {author} {\bibinfo {author} {\bibfnamefont {B.}~\bibnamefont
  {Murzaliev}}, \bibinfo {author} {\bibfnamefont {M.}~\bibnamefont {Titov}}, \
  and\ \bibinfo {author} {\bibfnamefont {M.~I.}\ \bibnamefont {Katsnelson}},\
  }\bibfield  {title} {\enquote {\bibinfo {title} {Diamagnetism of metallic
  nanoparticles as a result of strong spin-orbit interaction},}\ }\href
  {\doibase 10.1103/PhysRevB.100.075426} {\bibfield  {journal} {\bibinfo
  {journal} {Phys. Rev. B}\ }\textbf {\bibinfo {volume} {100}},\ \bibinfo
  {pages} {075426} (\bibinfo {year} {2019})}\BibitemShut {NoStop}%
\bibitem [{\citenamefont {G\'omez~Viloria}\ \emph {et~al.}(2021)\citenamefont
  {G\'omez~Viloria}, \citenamefont {Weick}, \citenamefont {Weinmann},\ and\
  \citenamefont {Jalabert}}]{gomez21_preprint}%
  \BibitemOpen
  \bibfield  {author} {\bibinfo {author} {\bibfnamefont {M.}~\bibnamefont
  {G\'omez~Viloria}}, \bibinfo {author} {\bibfnamefont {G.}~\bibnamefont
  {Weick}}, \bibinfo {author} {\bibfnamefont {D.}~\bibnamefont {Weinmann}}, \
  and\ \bibinfo {author} {\bibfnamefont {R.~A.}\ \bibnamefont {Jalabert}},\
  }\bibfield  {title} {\enquote {\bibinfo {title} {{Magnetic response of
  metallic nanoparticles: Geometric and weakly relativistic effects}},}\ }\href
  {\doibase 10.1103/PhysRevB.104.245428} {\bibfield  {journal} {\bibinfo
  {journal} {Phys. Rev. B}\ }\textbf {\bibinfo {volume} {104}},\ \bibinfo
  {pages} {245428} (\bibinfo {year} {2021})}\BibitemShut {NoStop}%
\bibitem [{\citenamefont {Hernando}\ \emph {et~al.}(2006)\citenamefont
  {Hernando}, \citenamefont {Crespo},\ and\ \citenamefont
  {Garc\'{\i}a}}]{herna06_PRL}%
  \BibitemOpen
  \bibfield  {author} {\bibinfo {author} {\bibfnamefont {A.}~\bibnamefont
  {Hernando}}, \bibinfo {author} {\bibfnamefont {P.}~\bibnamefont {Crespo}}, \
  and\ \bibinfo {author} {\bibfnamefont {M.~A.}\ \bibnamefont {Garc\'{\i}a}},\
  }\bibfield  {title} {\enquote {\bibinfo {title} {Origin of orbital
  ferromagnetism and giant magnetic anisotropy at the nanoscale},}\ }\href
  {\doibase 10.1103/PhysRevLett.96.057206} {\bibfield  {journal} {\bibinfo
  {journal} {Phys. Rev. Lett.}\ }\textbf {\bibinfo {volume} {96}},\ \bibinfo
  {pages} {057206} (\bibinfo {year} {2006})}\BibitemShut {NoStop}%
\bibitem [{\citenamefont {Politi}\ and\ \citenamefont
  {Pini}(2002)}]{polit02_PRB}%
  \BibitemOpen
  \bibfield  {author} {\bibinfo {author} {\bibfnamefont {P.}~\bibnamefont
  {Politi}}\ and\ \bibinfo {author} {\bibfnamefont {M.~G.}\ \bibnamefont
  {Pini}},\ }\bibfield  {title} {\enquote {\bibinfo {title} {Dipolar
  interaction between two-dimensional magnetic particles},}\ }\href {\doibase
  10.1103/PhysRevB.66.214414} {\bibfield  {journal} {\bibinfo  {journal} {Phys.
  Rev. B}\ }\textbf {\bibinfo {volume} {66}},\ \bibinfo {pages} {214414}
  (\bibinfo {year} {2002})}\BibitemShut {NoStop}%
\bibitem [{\citenamefont {Var\'on}\ \emph {et~al.}(2013)\citenamefont
  {Var\'on}, \citenamefont {Beleggia}, \citenamefont {Kasama}, \citenamefont
  {Harrison}, \citenamefont {Dunin-Borkowski}, \citenamefont {Puntes},\ and\
  \citenamefont {Frandsen}}]{varon13_SR}%
  \BibitemOpen
  \bibfield  {author} {\bibinfo {author} {\bibfnamefont {M.}~\bibnamefont
  {Var\'on}}, \bibinfo {author} {\bibfnamefont {M.}~\bibnamefont {Beleggia}},
  \bibinfo {author} {\bibfnamefont {T.}~\bibnamefont {Kasama}}, \bibinfo
  {author} {\bibfnamefont {R.~J.}\ \bibnamefont {Harrison}}, \bibinfo {author}
  {\bibfnamefont {R.~E.}\ \bibnamefont {Dunin-Borkowski}}, \bibinfo {author}
  {\bibfnamefont {V.~F.}\ \bibnamefont {Puntes}}, \ and\ \bibinfo {author}
  {\bibfnamefont {C.}~\bibnamefont {Frandsen}},\ }\bibfield  {title} {\enquote
  {\bibinfo {title} {Dipolar magnetism in ordered and disordered
  low-dimensional nanoparticle assemblies},}\ }\href {\doibase
  10.1038/srep01234} {\bibfield  {journal} {\bibinfo  {journal} {Sci. Rep.}\
  }\textbf {\bibinfo {volume} {3}},\ \bibinfo {pages} {1234} (\bibinfo {year}
  {2013})}\BibitemShut {NoStop}%
\bibitem [{\citenamefont {Alkadour}\ \emph {et~al.}(2017)\citenamefont
  {Alkadour}, \citenamefont {Mercer}, \citenamefont {Whitehead}, \citenamefont
  {Southern},\ and\ \citenamefont {van Lierop}}]{alkad17_PRB}%
  \BibitemOpen
  \bibfield  {author} {\bibinfo {author} {\bibfnamefont {B.}~\bibnamefont
  {Alkadour}}, \bibinfo {author} {\bibfnamefont {J.~I.}\ \bibnamefont
  {Mercer}}, \bibinfo {author} {\bibfnamefont {J.~P.}\ \bibnamefont
  {Whitehead}}, \bibinfo {author} {\bibfnamefont {B.~W.}\ \bibnamefont
  {Southern}}, \ and\ \bibinfo {author} {\bibfnamefont {J.}~\bibnamefont {van
  Lierop}},\ }\bibfield  {title} {\enquote {\bibinfo {title} {Dipolar
  ferromagnetism in three-dimensional superlattices of nanoparticles},}\ }\href
  {\doibase 10.1103/PhysRevB.95.214407} {\bibfield  {journal} {\bibinfo
  {journal} {Phys. Rev. B}\ }\textbf {\bibinfo {volume} {95}},\ \bibinfo
  {pages} {214407} (\bibinfo {year} {2017})}\BibitemShut {NoStop}%
\bibitem [{\citenamefont {Gallina}\ and\ \citenamefont
  {Pastor}(2020)}]{galli20_PRX}%
  \BibitemOpen
  \bibfield  {author} {\bibinfo {author} {\bibfnamefont {D.}~\bibnamefont
  {Gallina}}\ and\ \bibinfo {author} {\bibfnamefont {G.~M.}\ \bibnamefont
  {Pastor}},\ }\bibfield  {title} {\enquote {\bibinfo {title} {Disorder-induced
  transformation of the energy landscapes and magnetization dynamics in
  two-dimensional ensembles of dipole-coupled magnetic nanoparticles},}\ }\href
  {\doibase 10.1103/PhysRevX.10.021068} {\bibfield  {journal} {\bibinfo
  {journal} {Phys. Rev. X}\ }\textbf {\bibinfo {volume} {10}},\ \bibinfo
  {pages} {021068} (\bibinfo {year} {2020})}\BibitemShut {NoStop}%
\bibitem [{\citenamefont {Park}\ and\ \citenamefont
  {Stroud}(2004)}]{park04_PRB}%
  \BibitemOpen
  \bibfield  {author} {\bibinfo {author} {\bibfnamefont {S.~Y.}\ \bibnamefont
  {Park}}\ and\ \bibinfo {author} {\bibfnamefont {D.}~\bibnamefont {Stroud}},\
  }\bibfield  {title} {\enquote {\bibinfo {title} {Surface-plasmon dispersion
  relations in chains of metallic nanoparticles: An exact quasistatic
  calculation},}\ }\href {\doibase 10.1103/PhysRevB.69.125418} {\bibfield
  {journal} {\bibinfo  {journal} {Phys. Rev. B}\ }\textbf {\bibinfo {volume}
  {69}},\ \bibinfo {pages} {125418} (\bibinfo {year} {2004})}\BibitemShut
  {NoStop}%
\bibitem [{\citenamefont {Brandstetter-Kunc}\ \emph {et~al.}(2015)\citenamefont
  {Brandstetter-Kunc}, \citenamefont {Weick}, \citenamefont {Weinmann},\ and\
  \citenamefont {Jalabert}}]{brand15_PRB}%
  \BibitemOpen
  \bibfield  {author} {\bibinfo {author} {\bibfnamefont {A.}~\bibnamefont
  {Brandstetter-Kunc}}, \bibinfo {author} {\bibfnamefont {G.}~\bibnamefont
  {Weick}}, \bibinfo {author} {\bibfnamefont {D.}~\bibnamefont {Weinmann}}, \
  and\ \bibinfo {author} {\bibfnamefont {R.~A.}\ \bibnamefont {Jalabert}},\
  }\bibfield  {title} {\enquote {\bibinfo {title} {Decay of dark and bright
  plasmonic modes in a metallic nanoparticle dimer},}\ }\href {\doibase
  10.1103/PhysRevB.91.035431} {\bibfield  {journal} {\bibinfo  {journal} {Phys.
  Rev. B}\ }\textbf {\bibinfo {volume} {91}},\ \bibinfo {pages} {035431}
  (\bibinfo {year} {2015})},\ \bibinfo {note}
  {\href{https://link.aps.org/doi/10.1103/PhysRevB.92.199906}{Phys. Rev. B
  \textbf{92}, 199906(E) (2015)}}\BibitemShut {NoStop}%
\bibitem [{\citenamefont {Brandstetter-Kunc}\ \emph {et~al.}(2016)\citenamefont
  {Brandstetter-Kunc}, \citenamefont {Weick}, \citenamefont {Downing},
  \citenamefont {Weinmann},\ and\ \citenamefont {Jalabert}}]{brand16_PRB}%
  \BibitemOpen
  \bibfield  {author} {\bibinfo {author} {\bibfnamefont {A.}~\bibnamefont
  {Brandstetter-Kunc}}, \bibinfo {author} {\bibfnamefont {G.}~\bibnamefont
  {Weick}}, \bibinfo {author} {\bibfnamefont {C.~A.}\ \bibnamefont {Downing}},
  \bibinfo {author} {\bibfnamefont {D.}~\bibnamefont {Weinmann}}, \ and\
  \bibinfo {author} {\bibfnamefont {R.~A.}\ \bibnamefont {Jalabert}},\
  }\bibfield  {title} {\enquote {\bibinfo {title} {Nonradiative limitations to
  plasmon propagation in chains of metallic nanoparticles},}\ }\href {\doibase
  10.1103/PhysRevB.94.205432} {\bibfield  {journal} {\bibinfo  {journal} {Phys.
  Rev. B}\ }\textbf {\bibinfo {volume} {94}},\ \bibinfo {pages} {205432}
  (\bibinfo {year} {2016})}\BibitemShut {NoStop}%
\bibitem [{\citenamefont {Downing}\ \emph {et~al.}(2017)\citenamefont
  {Downing}, \citenamefont {Mariani},\ and\ \citenamefont
  {Weick}}]{downi17_PRB}%
  \BibitemOpen
  \bibfield  {author} {\bibinfo {author} {\bibfnamefont {C.~A.}\ \bibnamefont
  {Downing}}, \bibinfo {author} {\bibfnamefont {E.}~\bibnamefont {Mariani}}, \
  and\ \bibinfo {author} {\bibfnamefont {G.}~\bibnamefont {Weick}},\ }\bibfield
   {title} {\enquote {\bibinfo {title} {Radiative frequency shifts in
  nanoplasmonic dimers},}\ }\href {\doibase 10.1103/PhysRevB.96.155421}
  {\bibfield  {journal} {\bibinfo  {journal} {Phys. Rev. B}\ }\textbf {\bibinfo
  {volume} {96}},\ \bibinfo {pages} {155421} (\bibinfo {year}
  {2017})}\BibitemShut {NoStop}%
\bibitem [{\citenamefont {Jackson}(1962)}]{Jackson}%
  \BibitemOpen
  \bibfield  {author} {\bibinfo {author} {\bibfnamefont {J.~D.}\ \bibnamefont
  {Jackson}},\ }\href@noop {} {\emph {\bibinfo {title} {Classical
  Electrodynamics}}}\ (\bibinfo  {publisher} {Wiley \& Sons},\ \bibinfo
  {address} {New York},\ \bibinfo {year} {1962})\ \bibinfo {note}
  {{Sec.~5.6}}\BibitemShut {NoStop}%
\bibitem [{\citenamefont {N\'eel}(1949)}]{neel}%
  \BibitemOpen
  \bibfield  {author} {\bibinfo {author} {\bibfnamefont {L.}~\bibnamefont
  {N\'eel}},\ }\bibfield  {title} {\enquote {\bibinfo {title} {Th\'eorie du
  tra\^{i}nage magn\'etique des ferromagn\'etiques en grains fins avec
  applications aux terres cuites},}\ }\href@noop {} {\bibfield  {journal}
  {\bibinfo  {journal} {Ann. Géophys.}\ }\textbf {\bibinfo {volume} {5}},\
  \bibinfo {pages} {99} (\bibinfo {year} {1949})}\BibitemShut {NoStop}%
\bibitem [{\citenamefont {Frauendorf}\ \emph {et~al.}(1998)\citenamefont
  {Frauendorf}, \citenamefont {Kolomietz}, \citenamefont {Magner},\ and\
  \citenamefont {Sanzhur}}]{fraue98_PRB}%
  \BibitemOpen
  \bibfield  {author} {\bibinfo {author} {\bibfnamefont {S.}~\bibnamefont
  {Frauendorf}}, \bibinfo {author} {\bibfnamefont {V.~M.}\ \bibnamefont
  {Kolomietz}}, \bibinfo {author} {\bibfnamefont {A.~G.}\ \bibnamefont
  {Magner}}, \ and\ \bibinfo {author} {\bibfnamefont {A.~I.}\ \bibnamefont
  {Sanzhur}},\ }\bibfield  {title} {\enquote {\bibinfo {title} {Supershell
  structure of magnetic susceptibility},}\ }\href {\doibase
  10.1103/PhysRevB.58.5622} {\bibfield  {journal} {\bibinfo  {journal} {Phys.
  Rev. B}\ }\textbf {\bibinfo {volume} {58}},\ \bibinfo {pages} {5622}
  (\bibinfo {year} {1998})}\BibitemShut {NoStop}%
\bibitem [{\citenamefont {G\'omez~Viloria}(2021)}]{gomez_unpublished}%
  \BibitemOpen
  \bibfield  {author} {\bibinfo {author} {\bibfnamefont {M.}~\bibnamefont
  {G\'omez~Viloria}},\ }\href@noop {} {\bibfield  {journal} {\bibinfo
  {journal} {unpublished}\ } (\bibinfo {year} {2021})}\BibitemShut {NoStop}%
\bibitem [{\citenamefont {Richter}\ and\ \citenamefont
  {Mehlig}(1998)}]{richt98_EPL}%
  \BibitemOpen
  \bibfield  {author} {\bibinfo {author} {\bibfnamefont {K.}~\bibnamefont
  {Richter}}\ and\ \bibinfo {author} {\bibfnamefont {B.}~\bibnamefont
  {Mehlig}},\ }\bibfield  {title} {\enquote {\bibinfo {title} {Orbital
  magnetism of classically chaotic quantum systems},}\ }\href {\doibase
  10.1209/epl/i1998-00197-2} {\bibfield  {journal} {\bibinfo  {journal}
  {Europhys. Lett.}\ }\textbf {\bibinfo {volume} {41}},\ \bibinfo {pages} {587}
  (\bibinfo {year} {1998})}\BibitemShut {NoStop}%
\bibitem [{\citenamefont {Bonacchi}\ \emph {et~al.}(2021)\citenamefont
  {Bonacchi}, \citenamefont {Antonello}, \citenamefont {Dainese},\ and\
  \citenamefont {Maran}}]{Bonacchi}%
  \BibitemOpen
  \bibfield  {author} {\bibinfo {author} {\bibfnamefont {S.}~\bibnamefont
  {Bonacchi}}, \bibinfo {author} {\bibfnamefont {S.}~\bibnamefont {Antonello}},
  \bibinfo {author} {\bibfnamefont {T.}~\bibnamefont {Dainese}}, \ and\
  \bibinfo {author} {\bibfnamefont {F.}~\bibnamefont {Maran}},\ }\bibfield
  {title} {\enquote {\bibinfo {title} {Atomically precise metal nanoclusters:
  Novel building blocks for hierarchical structures},}\ }\href {\doibase
  https://doi.org/10.1002/chem.202003155} {\bibfield  {journal} {\bibinfo
  {journal} {Chem. Eur. J.}\ }\textbf {\bibinfo {volume} {27}},\ \bibinfo
  {pages} {30} (\bibinfo {year} {2021})}\BibitemShut {NoStop}%
\bibitem [{\citenamefont {Roda-Llordes}\ \emph {et~al.}(2021)\citenamefont
  {Roda-Llordes}, \citenamefont {Gonzalez-Ballestero}, \citenamefont {L\'opez},
  \citenamefont {Mart\'{\i}nez-P\'erez}, \citenamefont {Luis},\ and\
  \citenamefont {Romero-Isart}}]{roda21_PRB}%
  \BibitemOpen
  \bibfield  {author} {\bibinfo {author} {\bibfnamefont {M.}~\bibnamefont
  {Roda-Llordes}}, \bibinfo {author} {\bibfnamefont {C.}~\bibnamefont
  {Gonzalez-Ballestero}}, \bibinfo {author} {\bibfnamefont {A.~E.~Rubio}\
  \bibnamefont {L\'opez}}, \bibinfo {author} {\bibfnamefont {M.~J.}\
  \bibnamefont {Mart\'{\i}nez-P\'erez}}, \bibinfo {author} {\bibfnamefont
  {F.}~\bibnamefont {Luis}}, \ and\ \bibinfo {author} {\bibfnamefont
  {O.}~\bibnamefont {Romero-Isart}},\ }\bibfield  {title} {\enquote {\bibinfo
  {title} {Quantum size effects in the magnetic susceptibility of a metallic
  nanoparticle},}\ }\href {\doibase 10.1103/PhysRevB.104.L100407} {\bibfield
  {journal} {\bibinfo  {journal} {Phys. Rev. B}\ }\textbf {\bibinfo {volume}
  {104}},\ \bibinfo {pages} {L100407} (\bibinfo {year} {2021})}\BibitemShut
  {NoStop}%
\bibitem [{\citenamefont {Weick}\ \emph {et~al.}(2005)\citenamefont {Weick},
  \citenamefont {Molina}, \citenamefont {Weinmann},\ and\ \citenamefont
  {Jalabert}}]{weick05_PRB}%
  \BibitemOpen
  \bibfield  {author} {\bibinfo {author} {\bibfnamefont {G.}~\bibnamefont
  {Weick}}, \bibinfo {author} {\bibfnamefont {R.~A.}\ \bibnamefont {Molina}},
  \bibinfo {author} {\bibfnamefont {D.}~\bibnamefont {Weinmann}}, \ and\
  \bibinfo {author} {\bibfnamefont {R.~A.}\ \bibnamefont {Jalabert}},\
  }\bibfield  {title} {\enquote {\bibinfo {title} {Lifetime of the first and
  second collective excitations in metallic nanoparticles},}\ }\href {\doibase
  10.1103/PhysRevB.72.115410} {\bibfield  {journal} {\bibinfo  {journal} {Phys.
  Rev. B}\ }\textbf {\bibinfo {volume} {72}},\ \bibinfo {pages} {115410}
  (\bibinfo {year} {2005})}\BibitemShut {NoStop}%
\bibitem [{\citenamefont {Bouchiat}\ and\ \citenamefont
  {Montambaux}(1989)}]{bouch89_JP}%
  \BibitemOpen
  \bibfield  {author} {\bibinfo {author} {\bibfnamefont {H.}~\bibnamefont
  {Bouchiat}}\ and\ \bibinfo {author} {\bibfnamefont {G.}~\bibnamefont
  {Montambaux}},\ }\bibfield  {title} {\enquote {\bibinfo {title} {Persistent
  currents in mesoscopic rings: ensemble averages and half-flux-quantum
  periodicity},}\ }\href {\doibase 10.1051/jphys:0198900500180269500}
  {\bibfield  {journal} {\bibinfo  {journal} {J. Phys. (France)}\ }\textbf
  {\bibinfo {volume} {50}},\ \bibinfo {pages} {2695} (\bibinfo {year}
  {1989})}\BibitemShut {NoStop}%
\bibitem [{\citenamefont {Mermin}\ and\ \citenamefont
  {Wagner}(1966)}]{mermi66_PRL}%
  \BibitemOpen
  \bibfield  {author} {\bibinfo {author} {\bibfnamefont {N.~D.}\ \bibnamefont
  {Mermin}}\ and\ \bibinfo {author} {\bibfnamefont {H.}~\bibnamefont
  {Wagner}},\ }\bibfield  {title} {\enquote {\bibinfo {title} {Absence of
  ferromagnetism or antiferromagnetism in one- or two-dimensional isotropic
  {Heisenberg} models},}\ }\href {\doibase 10.1103/PhysRevLett.17.1133}
  {\bibfield  {journal} {\bibinfo  {journal} {Phys. Rev. Lett.}\ }\textbf
  {\bibinfo {volume} {17}},\ \bibinfo {pages} {1133} (\bibinfo {year}
  {1966})}\BibitemShut {NoStop}%
\end{thebibliography}%

\end{document}